\documentclass[12pt]{iopart}
\usepackage{graphics,graphicx,epsfig}
\usepackage{amssymb}
\usepackage{float}
\usepackage{color}
\usepackage{bbm}
\newcommand{\mod}{\textrm{~~mod }}
\newcommand{\dfrac}{\frac}
\newcommand{\text}{\rm}
\def\qed{$\Box$}

\newtheorem{prop}{Proposition}\def\PRO{\begin{prop}}\def\ORP{\end{prop}}
\newtheorem{coro}{Corollary}\def\COR{\begin{coro}}\def\ROC{\end{coro}}
\newtheorem{theo}{Theorem}\def\TH{\begin{theo}}\def\HT{\end{theo}}
\def\TH{\begin{theo}}\def\HT{\end{theo}}
\newtheorem{defi}[prop]{Definition}\def\DE{\begin{defi}}\def\ED{\end{defi}}
\newtheorem{lemme}[prop]{Lemma}\def\LE{\begin{lemme}}\def\EL{\end{lemme}}

\def\EQ#1{\begin{eqnarray}#1\end{eqnarray}}

\def\ket#1{{|}#1\rangle}
\def\bra#1{\langle#1{|}}
\def\op#1{\hat{#1}}

\def\op#1#2{|#1\rangle\langle#2|}
\def\braket#1#2{\langle #1 \mid  #2 \rangle}

\def\dm#1{\op{#1}{#1}}
\newcommand{\djj}{d\kern-0.4em\char"16\kern-0.1em}
\def\ora#1{\overrightarrow{#1}}

\makeindex
\begin{document}
 \title{Transformations between symmetric sets of quantum states }
\author{Vedran Dunjko}
\address{SUPA, Engineering and Physical Sciences, Heriot-Watt University, Edinburgh, UK}
\address{Division of Molecular Biology, Ru\djj er Bo\v{s}kovi\'{c} Institute, Bijeni\v{c}ka cesta 54, P.P. 180, 10002 Zagreb, Croatia}
 
\author{Erika Andersson}
\address{SUPA, Engineering and Physical Sciences, Heriot-Watt University, Edinburgh, UK}
\ead{\{vd51, e.andersson\}@hw.ac.uk}

\begin{abstract}

We investigate probabilistic transformations of quantum states from a `source' set to a `target' set of states. Such transforms have many applications. They can be used for tasks which include state-dependent cloning or quantum state discrimination, and as interfaces between systems whose information encodings  are not related by a unitary transform, such as continuous-variable systems and finite-dimensional systems. In a probabilistic transform, information may be lost or leaked, and we explain the concepts of leak and redundancy. % of probabilistic transforms. 
Following this, we show how the analysis of probabilistic transforms
significantly simplifies for symmetric source and target sets of states. In particular, we give a simple linear program which solves the task of finding optimal transforms, and a method of characterizing the introduced leak and redundancy in information-theoretic terms. Using the developed techniques, we analyse a class of  transforms which convert coherent states with information encoded in their relative phase to symmetric qubit states. Each of these sets of states on their own appears in many well studied quantum information protocols. Finally, we suggest an asymptotic realization based on \textit{quantum scissors}.
\end{abstract}
\maketitle
%\blue{\bf Had to comment the definition of eqnarray in iopart.cls in order for the compilation to go through.}
%\red{\bf I did not get Qcircuit to run, complaints about multiple definition of the command ket. So commented out the circuit diagram. Also has to redefine a few other commands (such as mod,  and dfrac).}

\section{Introduction}
Quantum information theory promises new and exciting ways to process information. 
Often the advantage a quantum protocol gives over a classical procedure lies in the fact that quantum states may be non-orthogonal. 
 Classical information may be encoded in non-orthogonal quantum states, as is the case for example in quantum key distribution. The classical information then cannot be fully extracted from the quantum state alone. Many physical systems are candidates for the realization of quantum processing, and often they perform well at distinct tasks. 
Therefore, future quantum devices may well be hybrid systems with interfaces linking the different parts, just as our classical information processing devices are today.
When classical information is encoded in a set of quantum states, %However, 
the information encodings of one system may be incompatible with the encodings of another in a sense which has no classical analogue: the `source' states may not be related to the corresponding `target' states by a fixed unitary transformation. 
%This occurs, \blue{for instance,} when the systems have different dimensionalities, such as continuous-variable (CV) quantum systems and finite-dimensional quantum systems. 
%\blue{\bf I am unsure this is not a bit misleading. There is no intrinsic problem with having distinct dimensionalities, as long as the relevant states occupy subspaces of equal dimensions. And even then it depends on the relative overlaps of the source and target sets, respectively... How about:}
This occurs, for instance, when we consider transforms between states of systems of distinct dimensionalities such as typical qubit states and coherent or sqeezed states of  continuous-variable systems.
%\blue{\bf Removed the acronym CV as it does not appear in the text later.}
When transferring information from one system to another where the encodings are incompatible in this sense, we must then either accept errors or resort to probabilistic scenarios where information may be lost, or leaked.
%Such events are 
This is important from an information-theoretic and cryptographic perspective. Information is no longer perfectly controlled by the emitting party. 
%\red{Probabilistic transforms from one set of quantum states to another can also be used for state-dependent probabilistic cloning and quantum state discrimination. In the former case, we are decreasing the overlap between the source states, and in the latter case, the target states are orthogonal.}

Transformations that take a `source' set of quantum states to a `target' set of quantum states, where the states in the two sets are not pairwise related by a single unitary transform, also have other applications. State-dependent quantum cloning is one example \cite{DuanGuo}.
Another related and well-studied family of such transforms solve the problem of amplifying coherent light, while keeping the coherent phase unaltered \cite{Grangier, Xiang, Zavatta, Croke, Jeffersampl}. This problem is very important in classical and quantum communication tasks over larger distances, and is usually resolved by generating approximations of amplified coherent states.
Optimal measurements for distinguishing between quantum states can also be seen as transforms taking some set of quantum states to mutually orthogonal states, followed by a measurement to distinguish the latter from each other. For so-called minimum-error measurements, pioneered by Holevo and Helstrom, the transforms are allowed to err, i.e. the declared output need not always be correct. Another tradition requires correctness but allows for result which declares that the transform (measurement) has failed, following the works of Ivanovi\'{c}, Dieks and Peres \cite{Ivanovic, Dieks,Peres}. %-- 
Such measurements are then called unambiguous \cite{chefles1998250,chefles-1998-31}.
%The study of 
%Transformations which take states from one set to another also have other applications. %has a long tradition. % and has been addressed with respect to various aspects. 
%In these works, 

Here we focus on %the unambiguous regime, where the transforms are 
unambiguous transforms
taking pure states to pure states. For this setting there exists a convenient framework based on the structures of the Gram matrices of `source' and `target' states, developed by Chefles, Jozsa and Winter \cite{chefles-2003,PhysRevA.65.052314}, %the main points of 
which we %presently 
will briefly present.
In these works,  the sets of source and targets states are general, and %the process of 
finding transforms for given sets of source and target states is complicated. %requires complicated optimization machinery.
However, %the case when 
it is known that for the problem of distinguishing quantum states which %\red {or transformed} %/analysed 
comprise a symmetric set, %has proven very convenient, in both regimes and in various settings, as it allows for  
a simpler treatment is possible %more elegant theory 
\cite{chefles1998250,Helstrom,Mixed}.  As we will show, this restriction simplifies the 
%induces a more elegant 
theory for %in the case of  
general probabilistic transforms as well. %, which operationally simplifies working with such states.
As an application of the theory we develop, we study the properties of converting %information encoded in the relative phase 
a set of coherent states to qubit states. This is an important %canonical 
example of an `interspecies' transform, as these two types of encodings frequently appear in %distinct
quantum information processing tasks.

\section{Preliminaries}
Our problem of interest is stated as follows: given two sets of pure states $A$ and $B$ (called `source' and `target' sets, respectively) of finite size $N$,
\begin{eqnarray}
&A=\{\ket{a_i}\}_{i=1}^{N} ; \ B=\{\ket{b_i}\}_{i=1}^{N},  & \nonumber
\end{eqnarray}
what are the properties of 
a transform $\mathcal{T}$, allowed by quantum mechanics, which performs $\mathcal{T}(\ket{a_i}) = \ket{b_i}$ for all $i$ perfectly with a certain probability?
%The most general case is given by a transform which is probabilistic -- it 
The transform can fail to produce the desired output state, or succeed, and these two possible outcomes are reported, i.e. the transform is heralded.

 In the most general case, the success probabilities may depend on which %the 
 source state we start from. %For this case, 
 We then have the following statement: %criterion:
\LE\label{Prob}
There exists a probabilistic transform taking each state $|a_i\rangle$ in $A$ to the state $|b_i\rangle$ in $B$, 
succeeding with the probabilities %$\{p_i\}_{i=1}^N$, 
$p_i$, for $i=1,\ldots , N$, %corresponding to the indices of the source states, 
iff there exist Gram matrices of kets $\Pi^s$ and $\Pi^f$ such that the equality
\EQ{\label{ProbCrit}
G_A = P^s \circ \Pi^s  \circ G_B + P^f \circ \Pi^f
}
holds, where 
\EQ{
P^s =  \biggl[ \sqrt{p_i p_j}    \biggr]_{i,j}\ and\ 
P^f =  \biggl[ \sqrt{(1-p_i) (1-p_j)}    \biggr]_{i,j},
}
and $G_A$ and $G_B$ are the Gram matrices of sets $A$ and $B$.
\EL
This is a special case of the Theorem 3 in \cite{chefles-2003}.
In the Lemma above, $\circ$ denotes the Hadamard (Shur, point-wise) matrix product, and the Gram matrix of the set of kets (or more generally vectors)  $A=\{\ket{a_i}\}_{i=0}^{N-1}$ is given by%:
\EQ{
G_A = \left[\braket{a_p}{a_q} \right]_{p,q},\ p,q = 0,\ldots ,N-1. \nonumber
}
The necessary and sufficient conditions for a matrix $M$ to be a Gram matrix of normalized kets (states) are that $i)$ $M$ is a positive-semidefinite matrix, and $ii)$ $M$ has unity across the main diagonal.

Such a quantum transform %(instrument) 
can be equivalently viewed, in the spirit of the Stinespring dilation, as %the following 
a unitary transform acting on an augmented Hilbert space,
\EQ{\label{ProbUnit}
U \ket{a_i}\ket{0}\ket{0} =\sqrt{p_i} \ket{b_i}\ket{\psi_i}\ket{0} + \sqrt{1-p_i} \ket{Fail}\ket{\phi_i}\ket{1}  \textup{\ for \ all \ }i,
}
where %the instrument is realized 
we learn whether the transform has succeeded or failed by measuring %out 
the third `indicator' register on the right-hand side of the expression. 
%The link between the two representations is given by noting 
One may show that the matrices $\Pi^s$ and $\Pi^f$ in expression (\ref{ProbCrit}) are the Gram matrices of the sets of kets $\{\ket{\psi_i}\}_i$ and $\{\ket{\phi_i}\}_i$, respectively.
%Note that in general, should the transform prescribed by the unitary in the expression 
If the transform in equation (\ref{ProbUnit}) succeeds, then the output registers contain the target state $\ket{b_i}$ but also a residual state $\ket{\psi_i}$ which may be correlated with the input state. From an information-theoretic perspective, this residual state may be seen as a leak of information, hence we call the set of states $\{\ket{\psi_i}\}_i$ the \textit{leak}.
If the states $\ket{\psi_i}$ are not correlated with the input state, which happens if and only if $\ket{\psi_i} = \ket{\psi_j}$ for all $i$ and $j$, then the transform is called \textit{leakless}.
Analogously, in case the transform fails, a fixed fail state is produced along with a residual state $\ket{\phi_i}$ is produced. The residual state may be correlated with the input state, and may be used to subsequently attempt to reconstruct the desired outcome. For this reason we call the set of states  $\{\ket{\phi_i}\}_i$ the \textit{the redundancy}. If all the states in the redundancy are identical, only then is the residual state uncorrelated to the input state, and the transform is called \textit{redundancy-free}.

If the success probabilities above do not depend on the source state ($p_i = p_j$ for all $i,j$), we call the transform \textit{uniform}. For this case the notion of the optimal transform can be naturally defined: a uniform probabilistic transform is optimal, if no other transform succeeds with a strictly grater probability.

%The cases of 
Deterministic and unitary transforms are easily seen to be %the 
special cases of %the 
probabilistic transforms. %, and the analogous criteria in terms of the Gram matrices are easily derived. 
For %the case of 
a deterministic transform, it holds that %the condition is given by setting 
$p_i=1$, in which case the criterion  $\label{ProbCrit}$ reads $G_A =  \Pi^s  \circ G_B$ (for some Gram matrix of states $\Pi^s$). For %the case of 
a unitary transform the complex matrix $\Pi^s$ is an outer product of a vector, containing roots of unity, with itself  (c.f. \cite{chefles-2003})~\footnote{This freedom in the complex phases reflects the fact that kets in general contain information about the physically irrelevant complex phase.}.
Throughout his paper, with $G_S$ we will denote the Gram matrix of the set of states $S$, and with $\lambda_{G_{S}}$ a vector comprising the eigenvalues of the matrix $G_S$. With $I$ we denote the identity matrix, and with  $\mathbf{1}$ we denote the matrix with unity at each entry.

\subsection{Example: uniform unambiguous discrimination of pure states \label{UDS}}
%The process of 
Unambiguous discrimination of states (UDS) identifies the input state from a pre-defined set of states, error free, but allows a `failure' option. It is equivalent to a probabilistic %(uniform) 
transform %with range in 
for which the states $|b_i\rangle$ are mutually orthogonal. %states 
  %and 
The criterion for the existence of such a transform %for a success probability $p$ 
is given by Lemma \ref{Prob%\ProbCrit
}. Since the Gram matrix of orthogonal states is the identity, and the Hadamard product of the identity and Gram matrix of states is the identity again, for the special case where the success probability $p$ is independent of the source state (uniform UDS), the existence condition simplifies to the inequality
\EQ{
G_A - p I \geq 0,
}
meaning that %which signifies the positive-semidefinitness of 
the matrix $G_A - p I$ is positive-semidefinite.
Since unitary basis change preserves operator positivity, and $G_A$ is positive-semidefinite, hence diagonalizable in an orthonormal basis, this %expression 
implies and is implied by
$$
p \leq \min \lambda_{G_{A}},
$$
where $\min \lambda_{G_{A}}$ denotes the smallest eigenvalue of the matrix $G_{A}.$
From this condition we easily capture a known result:  the optimal success probability of UDS is equal to the smallest eigenvalue of the Gram matrix $G_A$~\footnote{This result was proven using different techniques and stated in a different formalism in \cite{chefles1998250}.}. A consequence of this is another famous result:  a set of states may be unambiguously discriminated if and only if that set of states is linearly independent. 
The latter is clear as the spectrum of $G_A$ contains a zero element if and only if the set $A$ is linearly dependent.

%A less known consequence of this simple analysis (\todo{Where do I find this in literature} Erika: not sure) characterizes all optimal uniform probabilistic transforms:
From the fact that unambiguous state discrimination is possible iff a set of states if linearly independent, it is easy to see that
if a uniform probabilistic transform $\mathcal{T}$ is optimal, then the redundancy is a linearly dependent set of states. To prove this, assume that a uniform probabilistic transform $\mathcal{T}$ succeeds with probability $p$, and that the redundancy is linearly independent. Then, in the case of failure, one can run UDS on the redundancy, and %in the event 
if this %the UDS 
succeeds (with probability $p^\prime >0$, due to linear independence), the target state can still be generated from the outcome. %~\footnote{Such processes are often called \textit{measure and prepare} in literature.}. 
This overall procedure ($\mathcal{T}$ followed by UDS in case of failure) comprises a uniform probabilistic transform $\mathcal{T} ^\prime$ which performs the same task as $\mathcal{T}$ but succeeds with probability $p^\prime + p > p$. Hence $\mathcal{T}$ could not have been optimal.

\section{Transformations between symmetric sets of pure states}

As noted, the %restriction of quantum information tasks to the 
case when the sets of states in focus is symmetric %has been 
is of interest %both from purely theoretical reasons, and lately for practical reasons as well as  
since many %most 
quantum protocols  \cite{BB84,barbosa-2002, %} to more recent proposed quantum protocols \cite{
 10.1109/FOCS.2009.36,PhysRevA.74.022304, Leuchs}) work with symmetric quantum states. 
%Recall, 

A set of (pure) states $A=\{\ket{a_i}\}_{i=0}^{N-1}$ is \textit{symmetric} if there exists a fixed unitary $U$ with the property
$$
U \ket{a_i} = \ket{a_{(i+1)\,\!\! \! \!\! \mod N}}\ \textup{for\ all\ }i.
$$
The assumption that %of symmetricity on the 
source and target states are symmetric %immediately 
allows us to link probabilistic and uniform probabilistic transforms.  %-- 
In this case, any probabilistic transform can be `uniformized', as shown by the following lemma:
\LE\label{unif}(Uniformization)
If there exists a probabilistic transformation taking the states in $A$ to states in $B$, which succeeds with the probabilities $\{p_i\}_{i=1}^N$, where $A$ and $B$ are symmetric sets of states, then there exists a uniform probabilistic transform taking the states in $A$ to states in $B$ which succeeds with probability
$$
p = \dfrac{1}{N}\sum\limits_{i=1}^N p_i.
$$
\EL

The proof of this lemma is given in the Appendix.

Additional properties of uniform transforms with symmetric source and target states are rooted in the structural properties of Gram matrices of  sets of symmetric states: 
\LE\label{simcirc}
A Gram matrix of kets is a circulant matrix if and only if the corresponding set of kets is symmetric. 
\EL
Proof of this lemma is given in the Appendix.

%Recall,
A circulant matrix is a square matrix, defined by its first row, for which the $i^{th}$ row is the right-circular shift of the first row by $i-1$ positions. Circulant matrices frequently appear in signal processing, and have two convenient properties: $i)$ circulant matrices diagonalize when conjugated by the unitary discrete Fourier transform (DFT) matrix, and $ii)$ the discrete Fourier transform of the first row of the circulant matrix is a vector containing the eigenvalues of the circulant matrix \cite{matrixComp}.
%Recall, 
The discrete Fourier transform matrix of size $N$ is the Vandermonde matrix of the $N^{th}$ primitive roots of unity, given with 
$$
DFT = \left[ \exp \ \dfrac{-2 (p-1) (q-1) i \pi}{N}         \right]_{p,q}, ~~p=1,\ldots N,q=1,\ldots N 
$$
which, when scaled by the pre-factor $1/\sqrt{N}$ becomes unitary, and which we then denote $uDFT$.

The criterion for the existence of a uniform probabilistic transform taking states from $A$ to $B$, succeeding with probability $p$, is the existence of Gram matrices of states $\Pi^s$ and $\Pi^f$ such that the equation
\EQ{
G_A = p  \Pi^s  \circ G_B + (1-p)  \Pi^f \label{UnifProbCrit}
}
holds. This is a slight simplification of the more general %heterogeneous case the 
condition %for which is given 
in Lemma \ref{Prob}.

In general, probabilistic uniform transforms with symmetric source and target sets may have leak and redundancy which are not symmetric.
Nonetheless, the following lemma shows that such a transform has a variant with the same success probability where the leak and redundancy are symmetric:
\LE\label{symm} (Symmetrization)
If there exists a uniform probabilistic transform taking states from a symmetric set $A$ to a set of symmetric states $B$, succeeding with some probability $p$, then there exists a uniform probabilistic transform taking the states from $A$ to $B$, succeeding with probability $p$, where the leak and redundancy are symmetric. 
\EL

Proof of this lemma is given in the Appendix.

\subsection{Finding optimal uniform transforms \label{fot}} 
Both from a practical and theoretical point of view, when considering transforms from a source to a target set one is often most interested in the optimal transforms. Optimality is naturally defined only in the case of uniform transforms. 
However, by virtue of Lemma \ref{unif}, when transforms with symmetric source and target sets are concerned, if any kind of transform linking the source and target states exists, then so does a uniform transform.
In this sense, for transforms between symmetric sets, optimality can in principle always be defined as the optimality of the uniform transform~\footnote{One may be tempted to do the same for non-symmetric transforms. However  there exist non-uniform transforms which have non-symmetric source and/or target sets which succeed with non-zero probability for some states at least, for which no uniform transform exists (all uniform transforms fail with unit probability).}.
% \blue{\bf Comment to Erika: I think an example is the following: take a set of $N$ lin. indep. states, and define an UDS procedure, so the transform goes to orthogonal states. Now, add an additional state to the input set (the $N+1$st state) which is a linear combination of, say, the penultimate and the last state in the original input set. The new set is lin. dep. In this case, no uniform UDS exists, but you can define reciprocal vectors for the first N-2 states, and a vector which is orthogonal to the first N-2 states, yet overlaps with the space spanned by the last three states. Now, for this you can define a type of UDS, which 'recognizes', probabilistically, the first N-2 states, and the subspace of the last three. Declare the 'last three' outcome a fail as well. I think this is a non-uniform UDS, which clearly cannot be uniformized. Does this seem correct? \red{Seems so.}}}.
 
In general, given two sets of states $A$ and $B$, the quest for the optimal uniform transform taking the states in $A$ to states in $B$ reduces to the maximization of the success probability $p$ over the space of all positive-semidefinite matrices (of the appropriate size) $\Pi^s$ and $\Pi^f$ with unit diagonal, subject to the constraint given in expression (\ref{UnifProbCrit}).
The dimensionality of the search space is then quadratic in the number of states.
However, if source and target states are symmetric, as a consequence of Lemma \ref{symm}, we may assume that $\Pi^s$ and $\Pi^f$ are circulant as well. Then %Hence 
all the matrices appearing in expression (\ref{UnifProbCrit}) %above 
are circulant, as the Hadamard product of circulant matrices is also circulant.
Hence, they all diagonalize in the same basis, and the dimensionality of the search space reduces quadratically from $O(N^2)$ to $O(N)$, where $N$ is the number of states.

The problem of finding optimal uniform transforms which have symmetric source an target sets is resolved by the following canonical linear program:
% \begin{align}
%& \text{maximize}   & \ora{c}^\mathrm{T} .\ora{x}\\
%& \text{subject to} & M .\ora{x} \leq \ora{b} \\
%& \text{and} & \ora{x} \ge 0,
%\end{align}
 \begin{eqnarray*}
& \text{maximize}   & \ora{c}^\mathrm{T} .\ora{x}\\
& \text{subject~to~} & M .\ora{x} \leq \ora{b} \\
& \text{and} & \ora{x} \ge 0,
\end{eqnarray*}
where $\ora{c}^\mathrm{T} = \left[ 1, \ldots, 1 \right]$, $\ora{b} = \lambda_{G_A}$, and $M = DCM_{\lambda_{G_B}}$, which is a circulant matrix, where the $i^{th}$ \textit{column} is the vector $\lambda_{G_B}$ `downward' shifted by $i-1$ positions (the \textit{discrete convolution matrix} $DCM_{\lambda_{G_B}}$ of the vector $\lambda_{G_B}$).
The optimal success probability is given by
\EQ{p=\dfrac{\ora{c}^\mathrm{T}. \ora{x}}{N},}
where the dot `$.$' (e.g. $\ora{x}^{T}.\ora{y}$ or $M.\ora{x}$) denotes the standard matrix product. The vector of eigenvalues of the Gram matrix of the leak  of the optimal transform is given by  $\lambda_{\Pi^s}= \dfrac{1}{p} \ora{x}.$
%The coefficient vector $\ora{b}$ is the vector containing the eigenvalues of the Gram matrix of the set $A$ (denoted  $\lambda_{G_A}$).
%The matrix of the coefficents $M$ is a circulant matrix, where the $i^{th}$ \textit{column} is the `downward' shifted vector $\lambda_{G_B}$ by $i-1$ positions (\textit{the discrete convolution matrix} $DCM_{\lambda_{G_B}}$ of the vector $\lambda_{G_B}$, as we explain presently).
As both $G_A$ and $G_B$ are circulant matrices, the vectors of eigenvalues $\lambda_{G_A}$ and $\lambda_{G_B}$ are computed  by taking the discrete Fourier transform of the first row of $G_A$ and $G_B$, respectively.
 
In %For 
the remainder of this section we show that the linear program above solves the problem of finding optimal uniform transforms. The constraint (\ref{UnifProbCrit}) where all the matrices are circulant can be written in terms of the vectors of eigenvalues of the matrices appearing, as they all diagonalize in the same basis: %simultaneously diagonalize:
\EQ{
&\lambda_{G_A} = p \lambda_{\Pi^s  \circ G_B} + (1-p)\lambda_{ \Pi^f}.&  \label{lambda}
}
 Note that for the vector $\lambda_{ \Pi^f}$ to be a vector of eigenvalues of a circulant Gram matrix of states, it is sufficient and necessary that all its %the 
 entries are non-negative and sum up to $N$. 
Using the circular convolution Theorem it can be shown that 
\EQ{
\lambda_{\Pi^s  \circ G_B} = \lambda_{\Pi^s} \ast \lambda_{ G_B}, \label{prop1}
}
where $\ast$ represents the (normalized) discrete convolution (or discrete cross-correlation) of vectors defined al follows. 
If $\ora{x}$ and $\ora{y}$ are two vectors of size $N$, with corresponding entries $x_i$ and $y_i$ for $i = 0,\ldots,N-1$,
then $\ora{z}=\ora{x} \ast \ora{y}$ is a length $N$ (with components denoted $z_i$), defined component-wise by
\EQ{
z_i = \dfrac{1}{N}\sum_{j=0}^{N-1} x_{j} y_{ \left[ (N-j+i)\, \!\!\!\!\!\!\!\ \mod\, \!\! N \right]}. \label{prop2}
}
The discrete convolution of two vectors  can also be represented in terms of a matrix-vector product by using the discrete convolution matrix $DCM_{\ora{x}}$ of the vector $\ora{x}$ defined via its transpose: the transpose matrix $DCM_{\ora{x}}^T$ is a circulant matrix whose first row is the vector $\ora{x}$.
It holds that $\ora{x} \ast \ora{y} = DCM_{\ora{x}} . \ora{y} = DCM_{\ora{y}} . \ora{x} = \ora{x} \ast 
\ora{y} $.
Hence, the constraint (\ref{lambda}) is equivalent to
\EQ{
&\lambda_{G_A} = p\, DCM_{\lambda_{G_B}}\lambda_{\Pi^s}  + (1-p)\,\lambda_{ \Pi^f}& \label{const1} ,
}
which can be  shown to be equivalent to the inequality
\EQ{
&\lambda_{G_A} - p\, DCM_{\lambda_{G_B}}\lambda_{\Pi^s}  \geq 0, \label{const2}
}
where $\lambda_{\Pi^s}$ is a non-negative real vector, whose entries sum up to $N$. 
The inequality above is interpreted component-wise~\footnote{To prove this equivalence, note that (\ref{const1}) implies the constraint (\ref{const2}) as the eigenvalues of $\Pi^f$ have to be non-negative.
To see that the inverse holds as well, it suffices to show that if $\lambda_{\Pi^s}$ is a vector of positive components which sum up to $N$, then the entries of the vector $\lambda_{G_A} - p\, DCM_{\lambda_{G_B}}\lambda_{\Pi^s}  $ sum up to $(1-p) N$.
By construction, the entries of  $\lambda_{G_A}$ sum up to $N$. Recall that  $DCM_{\lambda_{G_B}}\lambda_{\Pi^s}$ is also a vector of eigenvalues of a Gram matrix of a symmetric set of kets. Hence its components also sum up to $N$. Hence, the components  of $\lambda_{G_A} - p\, DCM_{\lambda_{G_B}}\lambda_{\Pi^s}$ sum up to $N-p N = (1-p)N$, and we have shown the equivalence of constraints (\ref{lambda}), (\ref{const1}) and (\ref{const2}).}.
To obtain the linear program stated at the beginning of this section,  we note that if a vector $\ora{x}$ is a vector of length $N$ with non-negative entries $\{x_i\}$, which maximizes $s=\sum%\limits
_{i=1}^N x_i$ subject to the constraint
 \EQ{
   DCM_{\lambda_{G_B}} \ora{x}  \leq &\lambda_{G_A},
 }
 then $\lambda_{\Pi^s} = \dfrac{N}{s} \ora{x}$ allows for the maximal $p$ subject to constraint (\ref{const2}), and the maximum is reached at $p=\dfrac{s}{N}$.

\subsection{The geometric interpretation of the optimization procedure}
As we have shown, the search for the optimal probability of success $p_{opt}$ of a uniform transform which takes $N$ input states to $N$ output states, where both sets of states are symmetric, reduces to the following optimization problem:

$p_{opt}$ is the maximal $p$ subject to constraint
\EQ{
&\lambda_{G_A} = p\, DCM_{\lambda_{G_B}}\lambda_{\Pi^s}  + (1-p)\,\lambda_{ \Pi^f}&  .
}
where $\lambda_{\Pi^s}$ and $\lambda_{\Pi^f}$ are some non-negative real vectors, whose entries sum up to $N$.

We have also shown that the constraint above is equivalent to the inequality
\EQ{
&\lambda_{G_A} - p\, DCM_{\lambda_{G_B}}\lambda_{\Pi^s}  \geq 0 ,
}
where $\lambda_{\Pi^s}$ is a non-negative real vectors, whose entries sum up to $N$.

The search space defined by the constraint (\ref{const2}) is the space of all points embedded in an $N$ dimensional space whose coordinates sum up to $N$. This is a convex set, defined by the extreme points $\{e_i\}_{i=1}^{N}$, where $e_i$ is a vector with the number $N$ as the $i^{th}$ component, and zeroes elsewhere.
But then, by the linearity of matrix-vector multiplication, the set $$S=\left\lbrace DCM_{\lambda_{G_B}}\lambda_{\Pi^s} \vert \lambda_{\Pi^s}\geq 0, ||\lambda_{\Pi^s}||_1 = N  \right\rbrace$$
is a convex set as well. The norm $|| \cdot ||_{1}$ is defined as the sum of the absolute values of the entries of the vector in the argument.
It is easy to see that $S$ is the convex hull of the columns of the matrix $N \times DCM_{\lambda_{G_B}}$.
First, let us assume $ DCM_{\lambda_{G_B}}$ is non-singular, which is equivalent to saying that the set of target kets does not contain mutually orthogonal kets. 
Then it holds that the columns of the matrix  $N \times DCM_{\lambda_{G_B}}$ are also the extreme points of the set $S$.
The constraint (\ref{const2}) can then be written as 
\EQ{
&\lambda_{G_A} - p\, X \geq 0 \label{const3}
} 
where $X \in S$.

Let $T$ be the set defined as follows:
$$ 
T =conv(\{e_i\}_{i=1}^{N}),
$$
where $conv(A)$ denotes the convex hull of the set of points $A$.
$T$ is the convex set of all points which correspond to a symmetric set of $N$ states.
This is a regular $(N-1)$-simplex which we can embed in the vector space $\mathbbmss{R}^{N}.$
Clearly, the point $\lambda_{G_A}$ is an element of $T$, and $S$ is also a regular $N-1$-simplex, contained in $T$. 
Simplices $T$ and $S$ share their center at coordinates $(1, \ldots,1),$ and $S$ is in a scaled down, rotated copy of $T$. From this it can be shown that the rows of the matrix $ DCM_{\lambda_{G_B}}$ are the extreme points of the set $S$ even when the discrete convolution matrix is singular. The only exception is the degenerate case when all the entries of $ DCM_{\lambda_{G_B}}$ are equal, which corresponds to the case when the set of target states is an orthogonal basis.

\begin{figure} %[H] 
\centering
 \fbox{
   \includegraphics[scale=0.4]{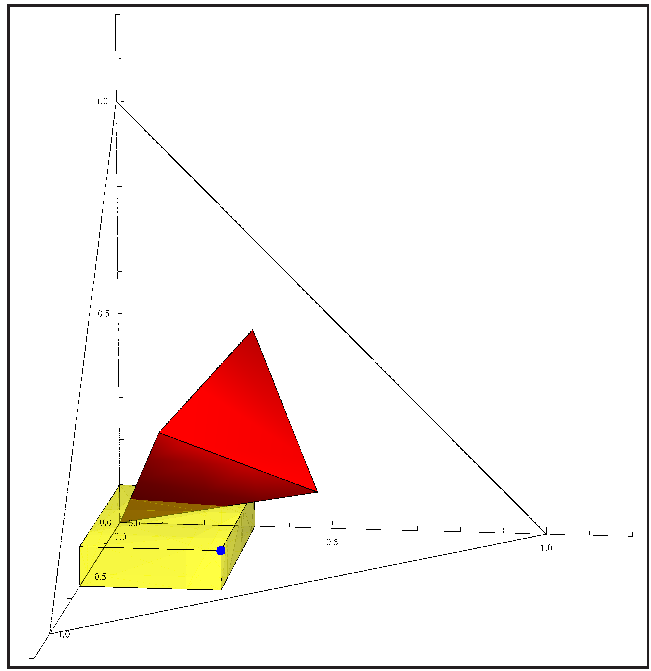}}

 \caption{\label{fig0} Illustration of the geometric interpretation of the solution existence criterion given in expression (\ref{const3}), for $N=3$.
 The the vector of eigenvalues of the Gram matrix of the source states is represented as a single point (blue in our illustration) which lies somewhere in the simplex defined by the extremal points $(0,0,1), (0,1,0),(1,0,0)$. This simplex is represented by the transparent triangle. The source states (represented by the blue point) uniquely define the orthotop $L$ -- a box, given in yellow. All points in the orthotope (and only those points) have the property that $i)$ %they have all non-negative components, 
 all their components are non-negative, and $ii)$ %then 
 any point in the orthotope when subtracted coordinate-wise from the blue point (defined by the source states) gives a point with non-negative components. The vector of eigenvalues of the Gram matrix of the target states defines the corresponding discrete convolution matrix, the columns of which are the extremal points of the search space $S$ defined at the beginning of this section. The space $S$ is, for $N=3$, a regular 2-simplex (moreover, an equilateral triangle), embedded in a 3-dimensional space, and is represented with the red triangle which lies within the transparent triangle representing all possible vectors of eigenvalues of the Gram matrix of a size three symmetric set of states. Together with the origin, the extremal points of $S$ define the 3-simplex  $p \times S = \{p \times x \vert x \in S\}$ which is represented by the entire red tetrahedron in the illustration. 
 Lemma \ref{geocrit} states that the necessary and sufficient criterion for the existence of a probabilistic transform taking the source to the target states is that the intersection between the red tetrahedron and the yellow box is not just the point of origin. The point of origin would correspond to a transform which succeeds with probability zero.
 Note also that clearly the intersection of the orthotope and the tetrahedron can be just the point of origin only if the orthotope has at least one dimension zero. This corresponds to the setting where source states are linearly dependent.
 }
\end{figure}

It is easy to see that, if $\lambda_{G_A}$ lies in $S$, then the constraint (\ref{const2}) can be satisfied for $p=1$, \emph{i.e.} there exists a deterministic transform from the set of states $A$ to the set of states $B$.
If this is not the case then the geometric interpretation of the constraint is as follows:
\LE\label{geocrit}
For a $0 < p \leq 1$ there exists a solution $X$ satisfying  the constraint (\ref{const3}) if the intersection between the simplex $p \times S = \{p \times \ora{x} \vert \ora{x} \in S\}$, embedded in $\mathbbmss{R}^{N}$, and the $(N)$ orthotope (hyperrectangle or box) $L$ defined by the opposite points  $(0,\ldots,0)$ and $\lambda_{G_A}$ is not the origin point alone.
\EL
The orthotope $L$ can be defined as %in the following way:
$$L = \left\{\ora{x} \in \mathbbmss{R}^{N} \vert \lambda_{G_A} - \ora{x}\geq 0 \right\},$$
which makes the validity of the geometric interpretation above obvious.
The geometric interpretation is illustrated in Figure \ref{fig0}, for the case $N=3.$  

Let us now consider a few special cases, for illustration purposes.
It is clear that if $S = T$ then for any set of input states the transform can be done deterministically, as $\lambda_{G_A} \in T.$
However, if  $S = T$, then $ DCM_{\lambda_{G_B}}$ has exactly one `1' %unity 
in each row and each column, hence the vector of eigenvalues $\lambda_{G_B}$ has one entry equal to $N$ and the rest is zero. This corresponds to the setting where the target set of states comprises exactly one state, and the deterministic transformation preforming this is the contraction to that particular state.

In the opposite scenario, %On the opposite side of the spectrum, 
$S$ may consist of a single point -- the point $(1,\ldots,1)$. In this case $ DCM_{\lambda_{G_B}}$ is a matrix containing just unities, and the corresponding set of target states is then orthogonal.
For there to exist a solution satisfying the constraint (\ref{const3}), by the geometric interpretation, the line $\{p (1,\ldots,1) \vert 0<p \leq  1 \}$ must intersect the orthotop $L$. This happens if and only if the extreme point $\lambda_{G_A}$ which defines the orthotope has all components non-zero.
This requirement implies that the input set of states is linearly independent. If we recall that a transformation with orthogonal target states is equivalent to unambiguous discrimination of input states, then we see we have recaptured a well-known result \footnote{Restricted, however, to symmetric sets of input states.}: a set of states can be unambiguously discriminated if and only if the set is linearly independent.

Finally we can use the geometric interpretation to give a new result, which we haven't addressed thus far:
If a uniform transformation with symmetric sets of input and output states is optimal, then the leak is linearly dependent.
 \begin{figure} %[H]
  \centering
 \fbox{
   \includegraphics[scale=0.4]{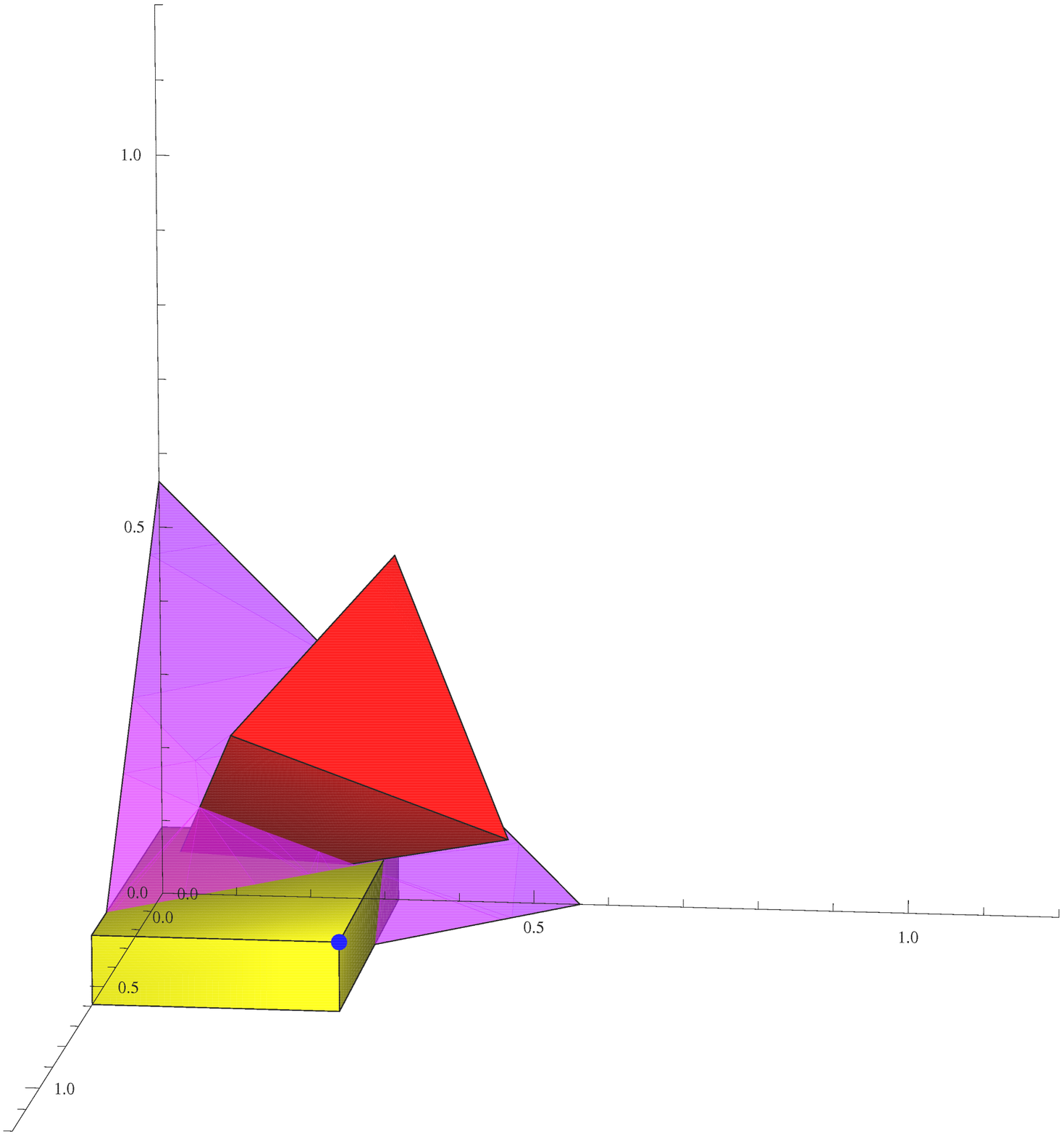}\hspace{0.5cm}
   \includegraphics[scale=0.25, ,viewport=0 0 550 650,clip]{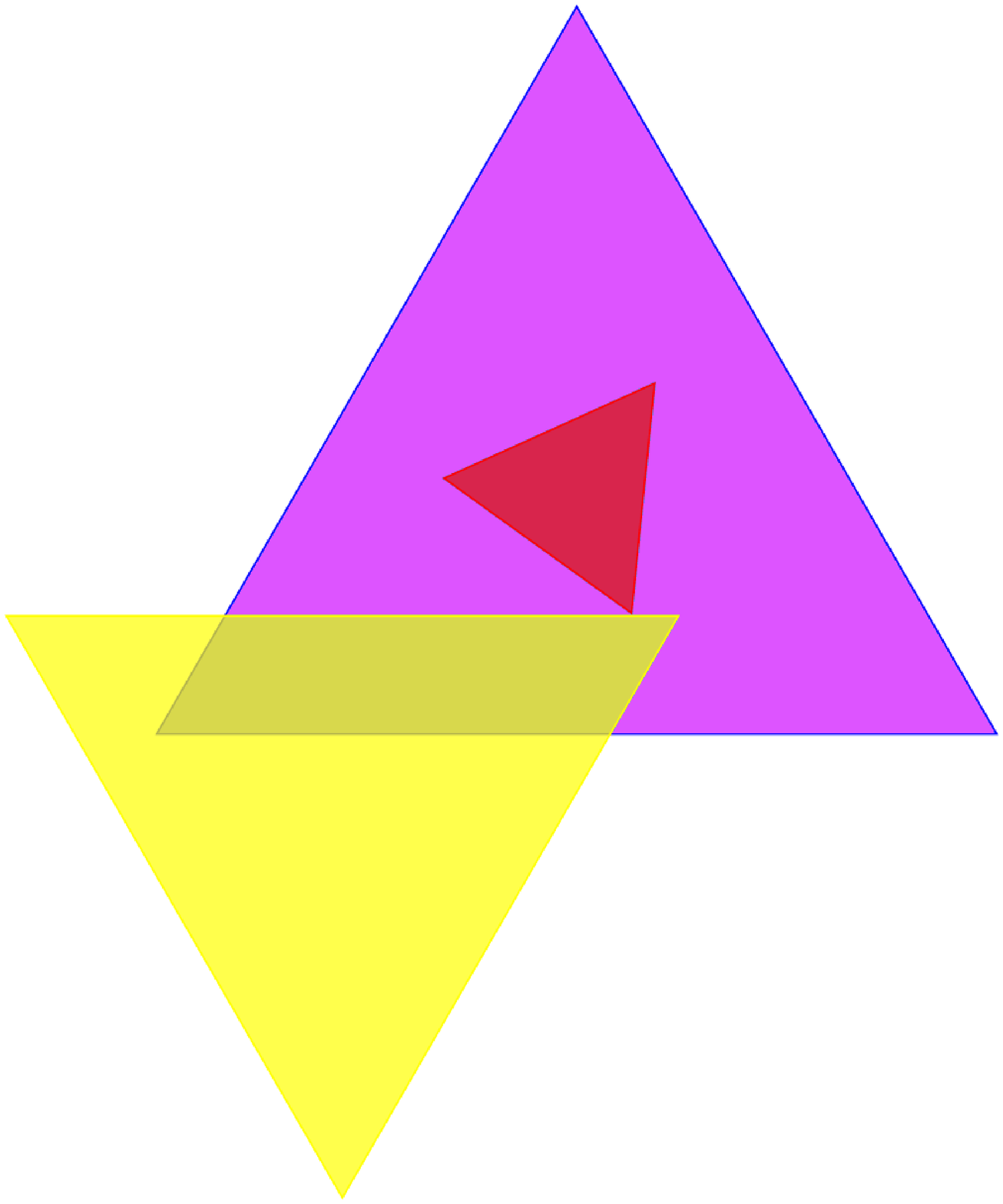}\hspace{0.5cm}
   }
 \caption{\label{fig8} Dynamic picture: The left-hand side illustration represents the setting where the optimal transform has been found. For a particular (maximal) $p$ the simplex which is the intersection of the (red) tetrahedron and the simplex $p \times T=conv((0,0,p), (0,p,0),(p,0,0))$ represented with the purple triangle touches the intersection of the yellow orthotope $L$ and the same simplex $p \times T$.
 The right-hand side image illustrates this event when the view is restricted only to the $N-1 = 2$-dimensional unique hyperplane $H_p$ containing the simplex $p \times T$. The intersection of the hyperplane $H_p$ and the whole red tetrahedron makes up the red triangle, the purple triangle is the convex set $p \times T=conv((0,0,p), (0,p,0),(p,0,0))$ in the same hyperplane.
 The yellow triangle corresponds to the intersection of the hyperplane $H_p$ and the extended orthotope $L^\prime$ where the edges of the original orthotope are allowed to extend to $-\infty$ each.
 This intersection is a regular 2-simplex again, with a fixed orientation with respect to  the simplex $p \times T$.
 In the dynamic picture, if we were to let $p$ slowly decrease from unity, we would witness the yellow triangle emerge from a single point, and grow until it touches the red triangle. The centre of the yellow triangle would slowly move towards the closest extremal point of $p \times T$ due to the change of its position in the barycentric coordinate system of the simplex $p \times T$ which changes as a function of $p$.
 }
\end{figure}
To show this we will adopt a dynamic picture as illustrated in Figure \ref{fig8}. 
Let $\lambda_{G_A} \not \in T$. What we seek is the largest $p$ such that the simplex $p \times S = \{p \times \ora{x} \vert \ora{x} \in S\}$ and the orthotope $L$ intersect. The simplex $p \times S$ clearly lies in the simplex $p \times T = \{p \times \ora{x} \vert \ora{x} \in T\}$, and the intersection will occur in this simplex.
As we slowly decrease $p$ the intersection between the simplex $p \times T$  and the orthotope $L$ `grows' while the simplex $p \times S$ slowly reduces in size. 
At one point, for some $p$, the intersection of  the simplex $p \times T$  and the orthotope $L$ touches the simplex $p \times S$, if there is a solution to the problem. Whenever this happens, the touching point is clearly on the face of the simplex $p \times S$, and not an interior point.

This means that the solution (the corresponding touching point in $S$) is a convex combination of at most $N-1$ rows of $ DCM_{\lambda_{G_B}}$, which in turn implies that $\lambda_{\Pi^s}$ has a zero component. Since $\lambda_{\Pi^s}$ is the vector of eigenvalues of the Gram matrix of the leak, this means the leak is linearly dependent.
If we now join this with the fact that optimal transforms have a linearly dependant redundancy, shown in section \ref{UDS}, we get the following statement:
\LE
If a uniform transformation with symmetric sets of input and output states is optimal, then the leak and the redundancy are linearly dependent.
\EL 

The inverse however, does not hold.

%\paragraph
\subsection*{Geometric characterization of the leak and the redundancy}
As we have shown, a uniform transform from symmetric to symmetric states succeeding with the probability $p$ can always be realized in such a way that the leak and the redundancy are symmetric sets of states (Lemma \ref{symm}). 
In this case, the leak and the redundancy can completely be characterized from the geometric picture.
Recall that, if the transform exists for a fixed $p$, then the intersection between the simplex  $p \times S$  and the orthotop $L$ is non-empty and this intersection is contained in the simplex $p \times T$. 
It is easy to see that the intersection $p\times F=L \cap p \times T$ is a convex set, more precisely, a bounded convex polytope.

Let $X$ be a solution, obeying the constraint (\ref{const3}). The vector $X$ completely characterizes the leak.
Recall, the vector $X$ is of the form $DCM_{\lambda_{G_B}} \lambda_{\Pi^s}$, where $\lambda_{\Pi^s}$ is the vector of eigenvalues of the leak set. Thus, the vector $X,$ viewed as a point in the simplex $p \times T$ embedded in the Euclidean space $\mathbbmss{R}^N$, is a convex combination of the rows of the matrix $DCM_{\lambda_{G_B}}$. The weights of this convex combination are the components of $\lambda_{\Pi^s}$. In other words, the representation of $X$ in the barycentric coordinates given by the extreme points of $p\times S$ (these points are the rows of the (scaled) matrix $DCM_{\lambda_{G_B}}$) gives exactly the vector of eigenvalues of the Gram matrix $\Pi^s$.
A barycentric coordinate system is a coordinate system in which a point's position is specified as the center of mass, or barycenter, of masses placed at the vertices of a simplex, in our case the simplex $p\times S$, which is the convex hull of the rows of the matrix $p \times DCM_{\lambda_{G_B}}$.

An analogous observation can be done for the redundancy -- $X$ represented in the barycentric coordinates of some of the the extreme points of $L \cap p \times T$ \footnote{The number of the extreme point of this polytope may be larger than $N+1$, but by Carath\'{e}odory's theorem, each point in this polytope can be represented as a convex combination of at most $N+1$ points.} will give us the structure of the redundancy.
While this relationship is more involved than in the case of the leak, and we leave it for further research, certain easy observations can be made for the optimal transform case.

As we noted, if the transform is optimal, then the solution point $X$ lies in the intersection of the faces of the polytope  $p\times F=L \cap p \times T$ and the simplex $p \times S$, \emph{i.e.} it is not in the interior of either.
If the dimensionality of the face which is involved in the contact of $p \times S$ is zero (a vertex) then every symmetrized optimal transform is always leakless.
Similarly, if the dimensionality of the face which is involved in the contact of $p \times F$ is zero, then it is redundancy-free.

If the contact involves faces of higher dimensionalities of $p \times F$ then essentially anything may happen, depending on the structure of the overlap.
In the example given in the right-hand side illustration of Figure \ref{fig8}, the contact point for the  simplex $p \times S$ (red) and the  simplex $p \times L^\prime \cap H_p$ (yellow) is a vertex of the simplex $p \times S$, and thus this transform is leakless. However, the contact point is interior of a 1-dimensional face of the yellow simplex, indicating that the vector of eigenvalues of the redundancy has two non-zero entries. Thus, the redundancy comprises at least two non-equal vectors. 

Note that the structure of the overlap depends on the relative orientations and positions of the polytope $p\times F$ and the simplex $p \times S$.
As we noted, the simplex $p \times S$ is just a scaled down and rotated simplex $p \times T$. The orientation of the polytope $p \times F$ is in a sense fixed with respect to the orientation of $p \times T$. To explain this,
consider the simplex $ L^\prime \cap p\times T^\prime$ where $L^\prime =\prod_{i=1}^{N} \left\langle -\infty , \lambda_{G_A}^{i} \right]$, $ \lambda_{G_A}^{i}$ being the $i^{th}$ component of the vector $ \lambda_{G_A}$ and the product is the Cartesian product.
We define $p T^\prime$ to be the hyperplane defined by the points $\{p \times e_i\}_{i=1}^N $.
The set $L^\prime$ is just the extended orthotope $L$ where the sides (1-faces) radiating from the point $\lambda_{G_A}$ are allowed to stretch to $-\infty$.
Then  $ L\prime \cap p\times T^\prime$  is the intersection of  $L^\prime \cap p\times T^\prime$ and the positive quadrant $\prod_{i=1}^{N} \left[ 0 , \infty \right\rangle.$
 $ L^\prime \cap p\times T^\prime$ is then a regular $N$-simplex, and if we translate it by moving the center to the point $(p, \ldots, p)$ we have a simplex which is a scaled, centrally mirrored copy of $p\times T$.
 In this sense, the orientation of  $p \times F$ (recall, $p \times F = \left( L^\prime \cap p\times T^\prime \right) \cap \prod_{i=1}^{N} \left[ 0 , \infty \right\rangle $) is fixed, relative to the orientation of  $p\times T$.

\subsection{Quantifying the leak and the redundancy}

Transformations between different types of quantum states become unavoidable when heterogeneous encodings are used for different aspects of quantum information tasks. In particular, such transform may be part %elements 
of a cryptographic protocol, in which case quantifying the leak and redundancy in information-theoretic terms becomes crucial.
For instance, one can imagine a simple two-party scheme in which party A, traditionally called Alice, wishes to send to party B, called Bob, information encoded in quantum states comprising the set of target states $B$.
However, Alice has at her disposal only quantum states from a set of quantum states $A$. So, Alice indeed does send her information encoded as states in $A$ to Bob, who performs an optimal probabilistic transform in order to obtain the target state $B$. For example, an ideal protocol may call for single-qubit states, but Alice can only generate pure states which approximate qubit states. It is then important for Alice to know what additional information Bob can obtain when transforming source states to target states~\footnote{Such approximations often appear in many proposals for realizations of quantum cryptographic protocols: polarization-encoded photons (which realize a qubit) are often approximated by polarized weak coherent pulses. In this case, almost without exception, a new security analysis is required.}.
As we have seen, such a transform is characterized by an expression of the form  $G_A = p  \Pi^s  \circ G_B + (1-p)  \Pi^f$
where the Gram matrices $G_A$ and $G_B$ fully characterize the source and target states (up to unitary equivalence), and  $\Pi^s$ and $\Pi^f$ characterize the \textit{leak}  and the \textit{redundancy}, that is, %-- 
the residual states when the transform succeeds and when it fails, respectively.
One way by which Alice may quantify the leak of information (embodied in the leak states) is by calculating the accessible information in this set of states. 
%This is upper bounded by the Holevo $\chi$ quantity.
If  $\{\rho_i \}_{i=0}^{N-1}$ is a set of %pure 
quantum states, then the accessible information $I_{acc}$ in this set of states is bounded above by the Holevo $\chi$ quantity,
$
I_{acc} \leq \chi (\rho_{AVG}) = S(\rho_{AVG}),
$
where $\rho_{AVG} = 1/N \sum_i \rho_i$ is the average state if each $\rho_i$ appears equally likely as a message, $S(\cdot)$ denotes the Von Neumann entropy,and the last equality holds if $\rho_i$ are pure.
If $\lambda=(\lambda_0, \ldots, \lambda_{N-1})$ is the vector of the eigenvalues of $\rho_{AVG}$, then the Von Neumann entropy can be expressed in terms of the Shannon entropy $H$ as
$
S(\rho_{AVG}) %= H(Q) 
= -\sum%\limits
_{i=0}^{N-1} \lambda_{i} \log \lambda_i. 
$
%where $Q$ is a discrete random variable with probability mass function defined by the probability vector $\lambda$.

If $A=\{\ket{a_i}\}_{i=1}^{N} $ is a set of kets (pure states), then using matrix algebra it can be shown that the non-zero eigenvalues of the matrix $G_A$ and the operator $\sum%\limits
_i \dm{a_i}$ are equal.
Hence, the upper bound on the accessible information in a set of states can be calculated as the  Shannon entropy of normalized eigenvalues of the Gram matrix of that set of states.
The optimization procedure we have presented, which finds the optimal success probability $p$, also finds the corresponding vector $\lambda_{\Pi^s}$. From this, %using which 
$\lambda_{\Pi^f}$ is easily computed, which are the eigenvalues of the Gram matrices of the leak and of the redundancy.
%As shown, 
From these eigenvalues it is then very simple to directly upper bound the accessible information in the leak and the redundancy.

\section{Application: From coherent states to qubit states \label{sect}}
Traditionally, for most applications of quantum information processing, the information %itself 
is encoded in qubit states.  However, it is also possible to use continuous-variable states, that is, %in recent years, the potential of using particular 
states of the quantum harmonic oscillator (e.g. coherent states). % for the same purpose has been extensively investigated. 
%In this section we investigate the feasibility of directly transforming one encoding to the other for two common types of information encodings in the quantum harmonic oscillator and qubit systems. 
In this section the source states will be a set of coherent states
\EQ{
A = \left\{\ket{a_k}= \ket{e^{i \theta_k} \alpha  } \right\}_{k=0,\ldots, N-1} \label{source}
}
where $\alpha$ is a real amplitude %of the coherent state 
and $\theta_k$ are their phases.
The target states are the qubit states in the Bloch sphere $XY$ plane,
\EQ{
B = \left\{\ket{b_k}=\dfrac{1}{\sqrt{2}}\left( \ket{0} + e^{i \theta_k} \ket{1} \right) \right\}_{k=0,\ldots, N-1}.\label{target}
}
By choosing the angles $\theta_k$ %to be defined 
as $\theta_k = 2 k \pi/N$ we obtain a very common family of encodings, which incidentally renders the sets $A$ and $B$ symmetric. 

\begin{figure} %[H] 
 \centering
   \includegraphics[scale=0.8]{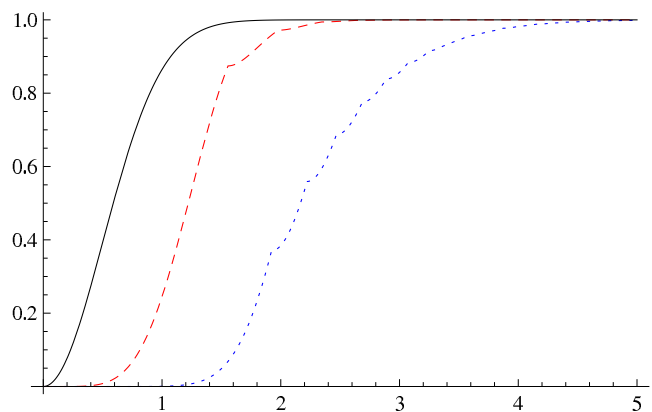}
 \caption{\label{fig1}
 (color online) 
 Comparison of the optimal success probability of unambiguous discrimination of  4 (red, dashed) and 8 (blue, dotted) states of a symmetric set of states, as a function of the real amplitude $\alpha.$ The black curve represents the optimal success probability of the coherent to qubit states transform, which is independent of the number of states.}
  \end{figure}

The problem we resolve is finding the optimal uniform transform taking the states in the set $A$ to those in $B$.
Initially, let us assume $N$ is even.
We may immediately note that the states in $A$ are linearly independent, so an unambiguous measure-and-prepare process will get us the desired transform succeeding with the success probability of an UDS procedure applied on the states in $A$. The optimal success probability of such a UDS procedure establishes a lower bound, and an upper bound is found by noting that if $N$ is even, then the desired probabilistic transform maps any two input states with relative phases differing by $\pi$ into orthogonal states. Hence, in particular this transform effectively performs unambiguous discrimination of the states $\ket{ \alpha  } $ and $|-\alpha\rangle$. %$\ket{\exp \left( i \pi \right) \alpha  }$. 
By using the results of section \ref{UDS}, the success probability of this UDS procedure (hence of the overall probabilistic transform) is upper bounded by $s_{bound} = 1 - \exp (-2 \alpha ^2)$.
This bound is always higher than the probability of unambiguous discrimination, except for the case of two states, where they coincide. The cases for 4 and 8 states are illustrated in Figure \ref{fig1}.

%For
In the remainder of this section we prove, constructively, that the upper bound can always be reached.  %and for this we employ additional mathematical machinery.
This is done by first obtaining results for the case $\alpha\le 1$, and then using these results, constructing transforms also for the case $\alpha>1$.

To begin, we introduce the notion of a multiprobabilistic transform, defined in \cite{chefles-2003}.
Multiprobablistic transforms are a generalization of probabilistic transforms, where there may be many different sets of targets states and with some probabilities an input state is transformed to a corresponding state in one of the target sets. %states. 
For our purposes, we shall define the uniform version of such transforms:
\DE
Let $S = \{\ket{s_i} \}_{i=1}^{n}$ be a set of source states and  $T^j = \left\{\ket{t_i^j} \right\}_{i=0}^{n}$  for $j=0,\ldots,k-1$ be a collection of possible target states.
A uniform multiprobabilistic transform $\mathcal{T}$ from the set $S$ to the sets in $\{T^j\}_j$, succeeding with the probability vector $(p_0, \ldots, p_{k-1})$, where $\sum\limits_{i=0}^{k-1} p_i  = 1$ and for all $i$ $p_i \geq 0$, performs 
$$\mathcal{T}(\ket{s_i}) =\ket{t_i^{j}} \ with\ probability\ p_{j} $$
for $i=1, \ldots, n$ and $j=0, \ldots ,k-1$.
\ED 
The set $T^0$ corresponding to success probability $p_0$ is reserved for the `fail outcome' states, analogous to the redundancy set of states in probabilistic transforms.

As a consequence of Theorem 3 in \cite{chefles-2003}, for fixed source set $S$ and target sets $\{T^j\}_{j=1}^{k-1}$ and a  probability vector $(p_0, \ldots, p_{k-1})$, such a uniform transform exists if and only if there exists a set of Gram matrices of states $\{\Pi^f, \Pi^1 ,\ldots ,\Pi^{k-1}\}$ such that the following equality holds:
\EQ{\label{uniMult}
G_S = p_0 \Pi^f + p_1 G_{T^1} \circ \Pi^1 + \cdots + p_{k-1} G_{T^{k-1}}\circ \Pi^{k-1},
}
where $G_S$ is the Gram matrix of the set $S$ and $G_{T^j}$ the Gram matrix of the set $T^j$ for all $j$.
We will call such a transform leakless if the matrices $\Pi^j = \mathbf{1}$ for all $j$ are matrices with all entries being the unity, and redundancy-free if the matrix $\Pi^f = \mathbf{1}$~\footnote{Note that $\mathbf{1}$ is a Gram matrix of any set of unit vectors which are all equal.}.

Let us now define a collection of sets of target states $B^j$ as
\EQ{
B^j = \{\ket{b_i}^{\otimes j}   \}_{i=0}^{N-1}, j=1,2,\ldots,N-1.
}
That is, the set $B^j$ comprises states which are $j$-fold copies of the elements of the (original target) set $B\equiv B^1$, which are the $XY$ plane qubit states. 
Then we have the following lemma, which holds specifically for the source and target states of interest:
\LE\label{lemmamulti}
Let the amplitude $\alpha$ of the states in the set $A$, defined in equation (\ref{source}), satisfy $0 < \alpha \leq 1$. Then there exists a uniform multiprobabilistic transform with the success probability vector $(p_0, \ldots, p_{N-1})$, which takes the states from the set $A$ to the collection of target states $\{B^j\}_{j=1}^{N-1}$ \emph{and} is redundancy-free and leakless. The failure probability %the probability 
$p_0$ of this transform is equal to $\exp (-2 \alpha ^2)$.
\EL
The proof of this Lemma is somewhat cumbersome and left for the Appendix.
The requirement that the transform be redundancy-free and leakless uniquely fixes the transform, up to the freedom in the choice of the realized fixed failure-outcome states.
% \red{\bf Specify what the fail state is, i.e. point out that it can be anything? Does this Lemma hold only for the sets $A$ an $B$ as specified in (13) and (14), or more generally? In the most general case, the fail probability can be equal to 0 (e.g. if the states in the set $A$ are orthogonal, but not only then).}

As a corollary of this Lemma we obtain the desired uniform probabilistic transform from coherent states in $A$, for $0<\alpha\le 1$, to the qubit states in $B$, and a characterization of the leak and redundancy for this optimal transform, as we now show. 
%\red{\bf Is the content of this corollary simply ``the transform in Lemma 6 is optimal?" Or is the transform in Lemma 6 not uniform? Point this out if so?} \blue{\bf Comment to Erika: The corollary simply says: if you can make a multiprobabilistic transform which generates copy-numbers of your desired targer state, then you can view the thing as a probabilistic (not 'multi') transform generating your target states. Lemma 6 is for uniform transforms only. Non-uniform multiprobabilistic transforms do not have a 'probability vector' but a sequence of probability matrices, which look like the success and fail probability matrices  $P^s$ and $P^f$ in expression (2) of Lemma 1.}

\COR \label{mult}
Let $A$ and $B$ be symmetric sets of an even number $N$ states, as defined at the beginning of this section, and let $0<\alpha\leq 1$. Then there exists a redundancy-free uniform probabilistic transform taking the states from $A$ to corresponding states in $B$ succeeding with probability $p_{succ} = 1-\exp(-2 \alpha^2)$. This transform is also optimal.
 \ROC

\noindent\textbf{Proof:\\}
Lemma \ref{lemmamulti} establishes the existence of a multiprobabilistic uniform transform from the set $A$ to the sets $\{B^j\}_{j=1}^{N-1}$, which is both redundancy-free and leakless,  when $0<\alpha \leq 1$. But then, by Theorem 3 in \cite{chefles-2003} there exists a probability vector $(p_0, \ldots, p_{N-1})$ such that the following equality holds:
\EQ{
G_A = p_0 \mathbf{1} + p_1 G_{B^1} + \cdots + p_{N-1} G_{B^{N-1}}. \label{one}
} 
Note, the expression above is the necessary and sufficient condition given in expression $(\ref{uniMult})$ for the existence of a uniform multiprobabilistic transform, which is now both redundancy-free, and leakless.

Since the Hadamard product is distributive, and by expression (\ref{bj}), this %the 
expression (\ref{one}) can be rewritten as
\EQ{
G_A = p_0 \mathbf{1} + (1-p_0) G_{B^1}\circ \left( \dfrac{p_1}{1-p_0} G_{B^0} + \cdots +     \dfrac{p_{N-1}}{1-p_0}  G_{B^{N-2}}\right),
}
with $ G_{B^0} = \mathbf{1}$.
Let us denote expression in the parenthesis in the equation above by $\Pi^s$,
 \EQ{
 \Pi^s =  \dfrac{p_1}{1-p_0} \mathbf{1} + \dfrac{p_2}{1-p_0} G_{B^1} + \cdots +     \dfrac{p_{N-1}}{1-p_0}  G_{B^{N-2}}.
 \label{structLeak}
 }
Note that $\Pi^s$ is a Gram matrix of states, as it is a convex combination of Gram matrices of states.
 So we have
 \EQ{
 G_A = p_0 \mathbf{1} + (1-p_0) G_{B^1}\circ \Pi^s.
 }
This expression is a sufficient criterion for the existence of a uniform probabilistic transform taking the defined coherent states to qubit states. Since the fail probability is $p_0= \exp(-2 \alpha^2)$, by the %considerations 
upper bound on the success probability derived at the beginning of this section, it is the lowest possible, and this transform is optimal. This transform is also redundancy-free, as the Gram matrix of the redundancy is $\mathbf{1}$, that is, a Gram matrix of a set comprising identical states. The leak of this transform is symmetric by lemma \ref{simcirc}, as the matrix $\Pi^s$ is a weighted sum of circulant matrices (see expression (\ref{structLeak})), hence circulant itself.   
%\red{\bf explain why?}.
\qed
 
By investigating the expression (\ref{structLeak}), we can construct the leak states of this transform explicitly. The leak state $\ket{\psi_i},$ corresponding to the input state $\ket{a_i}$ can, up to unitary equivalence, be written as
\EQ{\label{leakyst}
\ket{\psi_i} = \sum\limits_{j=0}^{N-2} \sqrt{\dfrac{p_{j+1}}{1-p_0}} \ket{b_i}^{\otimes j} \otimes \ket{0}^{\otimes N-2-j} \otimes \ket{j} \label{leak}
}
where the states of the last register (the \emph{indicator} register) are orthogonal for differing labels, and we define for any state $\ket{\eta}$, the zeroth tensoral power $\ket{\eta}^{\otimes 0} \equiv 1$ (the unity of the field underlying the Hilbert space, i.e. the number one). 
%\red{\bf meaning unit operator?}.
These `leaky' states are superpositions of varying numbers of copies (from zero to $N-2$) of the target state $\ket{b_i}$, all living in orthogonal subspaces of a larger Hilbert space (%as defined by 
due to the orthogonality of the indicator register states).

%However, 
We will now %wish to 
prove the existence of an optimal transform for any amplitude, also $\alpha>1$. To do this we first note that coherent states can be `split' into multimode states of a lower amplitude, i.e. there exists an isometry performing $U \ket{e^{i \phi} \alpha} = \bigotimes_{k=0}^{M-1} \ket{e^{i \phi} \beta_k}, \forall\ \phi$, as long as $\alpha^2 = \sum_{k=0}^{M-1} \beta_k^2.$ We %omit the proof, but 
note that %for instance, 
in quantum optics, this transform can be implemented by using balanced %perfect weighted 
beamsplitters and phase shifters.
Assume that we are given a set of coherent symmetric states $A$, as defined in equation (\ref{source}) with $\theta_k=2\pi k/N$, 
%which were defined as
%\EQ{
%A = \left\{\ket{a_i}= \ket{e^{i \theta_k} \alpha} \right\}_{k=0,\ldots, N-1},
%} 
%with the angles $\theta_k$ 
%as defined at the beginning of this section, 
of amplitude %greater than 
$\alpha>1$.
Each of these states in $A$ can be deterministically taken to the state $\bigotimes_{k=0}^{M-1} \ket{e^{i \theta_k} \beta}$ by `splitting' the coherent state into $M$ modes, where
 $\beta = \dfrac{\alpha}{\lfloor \alpha \rfloor +1}$ and  $M = \left( \lfloor \alpha \rfloor +1 \right)^2$.
Now we have that $\beta \leq 1$ and $\alpha^2 =M \beta^2 ,$ where $M$ is a non-negative integer.
By the Corollary $\ref{mult}$ we have that each subsystem state $\ket{e^{i \theta_k} \beta}$ can be individually transformed to the corresponding qubit state in the set $B$ with probability $\exp (-2 \beta^2).$
Note that, if only one of the individual transforms performed on the states $\ket{e^{i \theta_k} \beta}$ succeeds, then we have succeeded in generating exactly one copy of the target state from the source state $\ket{e^{i \theta_k} \alpha}$.
The probability of the transform failing on all $M$ copies is
%\EQ{
$\exp (-2 \beta^2)^M = \exp (-2 \alpha^2)$.
%\nonumber
%}
Hence, we have the following Theorem.
\TH\label{maintheo}
Let $A$ and $B$ be symmetric sets of an even number $N$ states, as defined in equation (\ref{source}) with $\theta_k=2\pi k/N$, %at the beginning of this section, 
and let $\alpha > 0$. Then there exists a redundancy-free uniform probabilistic transform taking the states in $A$ to the corresponding states in $B$,
 succeeding with probability $p_{succ} = 1-\exp(-2 \alpha^2)$. This transform is optimal.
\HT

The leak of this overall transform will in general comprise multimode states, which in some modes contain a fixed state (the modes where the probabilistic transform failed), and in some modes the target qubit and the individual transform leak of the form given in expression $(\ref{leak})$.
In contrast to unambiguous discrimination procedures for symmetric coherent states, the success probability of these optimal transforms generating qubit states does not depend on the number of states. 
In this analysis, we have assumed that the number of possible input states is even. As the success probability does not depend on the (even) number of states, the probabilistic transform can be done with the same success probability even when the number of states is $N$ for an odd $N$. To see this, simply consider the transform which works for $2N$ states. The initial odd numbered symmetric states will be an interlaced subset of the extended set.
However, here we do not have the validity of the upper bound any more, and it is not clear this success probability is optimal. While we do not offer a proof that the same bound holds for odd numbered states, evidence from performed numerical testing confirms this hypothesis.

An interesting aspect of the presented transform is that the success probability %which can be realized 
does not depend on the number of source and target states. Therefore it is possible that the same success probability may be reached when we consider the limit of an infinite number of states, $N \rightarrow \infty$. However, in the proofs of lemmas in this analysis, the finiteness of $N$ is used, so proving this extension to the limit may be non-trivial.
In the following section, we will however present a proposal for the realization of the presented transform, which does not assume a finite number of states, but achieves optimality in an asymptotic limit only. 

%\red{\bf We can take $N\rightarrow \infty$ without affecting the success probability -- say something about this?} \blue{\bf Comment to Erika: Actually, in the proof of the relevant Lemma 6. the finite dimensionality of the systems is assumed throughout. At the face of it, I cannot claim the proof extends to the limit $N \rightarrow \infty$ trivially. However, the construction of the transform using the quantum scissors we give below is independent of the angles and works with a continuum of angles. But there, we need to allow the splitting procedure to go to infinity. I am uncomfortable stating that it indeed holds for the infinity case. We can embed the instance of each transform for every even N into the infinite dimensional space, and claim that this sequence converges. I guess the infinite space (of the inf-dim Stinespring unitaries realizing the transform) will also be a Hilbert space (specifically, complete so every Cauchy sequence converges), but one should actually prove that it is a Cauchy sequence given the natural metric (whichever that is) of the inf-dim operator space. This may take a bit of work, if it is interesting enough? \red{\bf The proof should work for any finite $N$ though, no matter how large, always giving the same success probability for any $\alpha$? Should not the resulting `success probability sequence' then converge? No need to go into too much detail if you think a rigorous treatment is not trivial.}}

 \subsection{Transforming coherent to qubit states using optical state truncation}
 
 After these results on the existence of optimal transforms, we will look at practical ways of implementing such transforms.
%From the structure of coherent states it is clear that 
A straightforward way of (sub-optimally) generating the desired qubit states from the source coherent states %may be achieved 
is through optical state truncation (OST)~\cite{PhysRevLett.81.1604} or `quantum scissors', as we will now describe. 
For a single mode state, such as a coherent state, OST is the probabilistic and heralded projection of the input state to a finite subspace (as defined by a selection of a number of Fock states), followed by renormalization of the state vector. %Optical setups which realize, probabilistically and in a heralded way, optical state truncation are generically called \textit{quantum scissors} \cite{PhysRevLett.81.1604,
%Optical state truncation 
OST has been realized using a linear optical network~\cite{sciss2}. 
 In this section we will focus on %optical state 
 truncation to the subspace of the first two Fock states. %Formally, 
 Given the input state expanded in the number basis,
 \EQ{
 \ket{\psi}= \sum\limits_{i=0}^{\infty} c_i \ket{i}, \nonumber
 }
 where %the squares of the norms of the coefficients $c_i$ sum up to unity,  
$\sum_i|c_i|^2=1$,
 %the optical state truncation process 
 OST is characterized by the POVM (POM) elements 
 \EQ{
 \Pi_s = \dm{0}+\dm{1}, \ \
 \Pi_f = I - \Pi_s
 }
 and, upon success, produces the state
 \EQ{
 \ket{\psi_{trunc}} = \mathcal{N} \left( c_0 \ket{0} + c_1\ket{1} \right) \nonumber
}
 where the normalization factor is $\mathcal{N} = \left(|c_0|^2 + |c_1|^2 \right)^{-1/2} $.
 If we now consider the input state to be a state from our source set of $N$ coherent states,
 \EQ{
 \ket{a_j} :=  \ket{e^{\theta_j i} \alpha  } = e^{{-\alpha^2}/{2}} \sum\limits_{k=0}^{\infty} \dfrac{\alpha^k e^{k \theta_j i}}{\sqrt{k}} \ket{k},
 } 
 we see that the output state, after successful %optical state truncation (OST), 
 OST, which occurs with probability $p^{OST} = e^{-\alpha^2} (1+\alpha^2)$, is
 \EQ{
 \ket{{a_{j}}^{OST}} = \dfrac{1}{\sqrt{1+\alpha^2}} \left(\ket{0} + \alpha e^{i \theta_j} \ket{1}\right). \label{OST}
 }
 If $\alpha = 1,$ this transform %, for $\alpha =1$  performs exactly the discussed transform taking coherent to qubit states.
produces exactly the desired target qubit states.
 
This realisation does not, however, give the optimal success probability.  The success probability of this transform for $\alpha =1$ is approximately 0.735, which is less than the optimal value of approximately 0.864.
 The success probability of optical truncation to the vacuum and single photon subspace  approaches unity more than exponentially quickly as the amplitude tends to zero. For $\alpha \not= 1$, the truncation will not produce the targeted qubit state, due to an uneven distribution of the weights between the $\ket{0}$ and $\ket{1}$ states. Re-weighting of the amplitudes can, however, also be achieved probabilistically, so now we consider the performance of the coherent to qubit transform realized by state truncation, followed by redistribution of the weights, for amplitudes $\alpha< 1$.
 
The redistribution of weights may optimally be done by applying a POVM defined by the positive elements
 \EQ{
 P_f = \gamma \dm{0}, \ \
 P_s = I - P_f,
 }
 where $\gamma = 1 - \alpha^2$.
 These transforms fall into a class we call \emph{umbrella transforms}.
 The success rate (the probability of outcome associated with $P_s$) of this transform is 
 $p_{umb}= {2 \alpha^2}/({1+\alpha^2}),$
 hence the overall success probability of optical truncation followed by an umbrella transform for weight redistribution is
 \EQ{
 p_{overall}= p_{umb}p^{OST}= \dfrac{2 \alpha^2}{1+\alpha^2}e^{-\alpha^2} (1+\alpha^2) = 2 \alpha^2 e^{-\alpha^2}. \nonumber
 }
 This value is always below the success probability of the optimal transform as
 the quotient $p_{opt}/p_{overall}$ is equal to
 \EQ{
 \dfrac{p_{opt}}{p_{overall}}= \dfrac{\sinh (\alpha^2)}{\alpha^2},
 }
 which is always greater than 1 on the interval of interest, approaching unity when $\alpha\rightarrow 0$.
 
 %
 % \begin{figure}[h] \centering
 % \fbox{
 %   %\includegraphics[scale=0.8]{ill4.pdf} \hspace{0.5cm}
 %    % \includegraphics[scale=0.8]{ill5.pdf}
 %   \epsfig{file=ill4,width=0.9\linewidth,clip=} \hspace{0.5cm}
 %   \epsfig{file=ill5,width=0.9\linewidth,clip=}}
 %    
 % \caption{\label{fig2} Comparison of the success probability of the optimal coherent to qubit states transform and the transform realized by quantum scissors, followed by relative weight normalization between the vacuum and non-vacuum components. In the graph on the left-hand side, the full black curve and the red dashed curve represent the success probabilities of the optimal transform, and the transform based on quantum scissors, respectively. The right-hand graph gives the quotient between the two success probabilities with the black curve, and the dashed red line emphasizes the unit value. }
 % \end{figure}

 \subsection{Asymptotic optimality through beamsplitting}
 
 While the optimal transform of coherent to qubit states cannot be realized by OST %optical state truncation 
 followed by an umbrella transform to redistribute relative weights, it is evident that this transform performs better and better as the amplitude is reduced.
 It is natural to check whether a beamsplitting pre-procedure, analogous to the one used to prove the optimality Theorem \ref{maintheo} in the $\alpha>1$,  may be used to boost the overall success probability. % of the coherent to qubit state transform using optical truncation and an umbrella transform. 

The procedure goes as follows: the input state of real amplitude $\alpha$ is `beamsplitted' into $M$ modes of amplitude $\alpha/\sqrt{M}$ with the same complex phase as the initial beam (as was done in the proof of Theorem \ref{maintheo}). Then OST is applied to each of the beams, %the optical truncation is applied, 
and if an individual OST succeeds, %succeeded, 
an umbrella transform is applied to re-weigh the vacuum and $\ket{1}$ components.
The overall procedure succeeds if, for at least one of the split off beams, both the truncation and the umbrella transform are successful.
 
\begin{figure} %[h]
   \centering
  
    \includegraphics[scale=0.8]{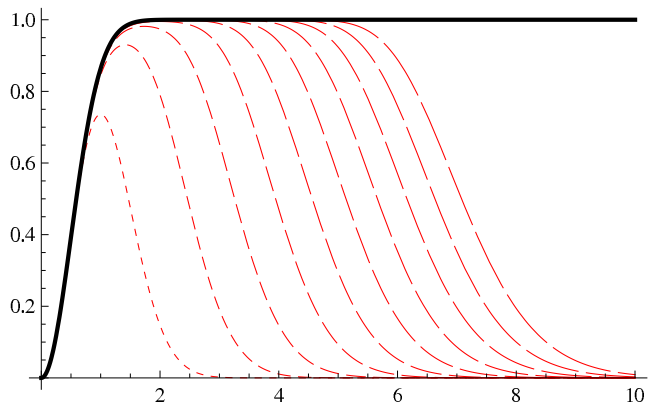}
  \caption{\label{fig6} (color online) Comparison of the success probability of the optimal coherent to qubit states transform and the transform realized by beamsplitting into $M$ beams of equal real amplitudes, followed by quantum scissors, followed by relative weight normalization between the vacuum and non-vacuum components on each of the weaker beams. 
  The $x$ axis gives the input amplitude $\alpha$ and $y$ the success probabilities.
  The full (black) curve is the success probability of the optimal transform, and the (red) dashed curves the success probabilities of the beamsplitter-assisted quantum scissors strategies for $M = 1 \ldots 10.$ The longer-dashed curves correspond to larger parameter $M$. }
  \end{figure}

 As we have shown, for a real amplitude $\alpha$, a re-weighted OST %quantum scissors transform which 
 produces the corresponding qubit state succeeds with probability
  $
  p_{overall} = 2 \alpha^2 e^{-\alpha^2}.
  $
 Then, the success probability of the strategy where the input beam has been split into $M$ beams is given by %with the following expression:
  \EQ{
   p_{overall,M} =1-\left(1- 2 \dfrac{\alpha^2}{M} e^{-\alpha^2/M}\right)^M.
  }
  In the asymptotic case of infinitele many `splits', the failure probability becomes
  \EQ{
  p_{overall,\infty}= \lim\limits_{M\rightarrow \infty} \left(1- 2 \dfrac{\alpha^2}{M} e^{-\alpha^2/M}\right)^M = e^{-2\alpha^2}, \nonumber
  }
  which is equal to the  failure probability of the optimal transform.
 % \red{{\bf Do we want to reinsert this? Ideally, figures should be referred to in the text. Was there a reference in the text to fig 1? I can't find anything commented out here.} 
 The graph in Figure \ref{fig6} %\blue{NoGraph??} 
  compares the success probabilities of the optimal transform and the beamsplitter-assisted strategies for various numbers of splits $M$.
  This procedure can then arbitrarily well approach the optimal success probability. % and %unlike the optimal transform analysed in the previous section, %this approach 
 It is suitable for experimental realizations, as both quantum scissoring and the weight redistribution using umbrella transforms %have been 
 may be realized experimentally.

\section{Conclusions}
In this work we have addressed probabilistic transforms taking states from a `source' to a `target' set of quantum states, with emphasis on the case where these sets are symmetric. 
Such transforms can for example serve as interfaces between continuous-variable and finite-dimensional quantum systems. 
%as a means of converting of one encoding of classical information to an incompatible (not unitarily equivalent) encoding. Such transforms may play a vital role in, for instance, composite quantum information protocols comprising subroutines which rely on mutually incompatible information encodings  where they serve as an information interface.
State-dependent cloning and quantum state discrimination are also special cases of probabilistic transforms.

We have emphasised that in %such a setting, 
a probabilistic transform, information may be lost and leaked, which may have impact on the protocol efficiency or security. For this purpose we introduced the concepts of the leak and redundancy of a probabilistic transform.
We have demonstrated how symmetric source and targets sets, which arise naturally in many quantum information applications, allows for a much simpler theory.
In particular, we derived a linear program which finds optimal uniform probabilistic transforms in this symmetric setting. This constitutes a significant simplification over %general 
optimization techniques which must be employed in %most 
more general cases, and the dimensionality of the search space is reduced quadratically in the number of states considered. The presented method also allows for a simple characterization of the aforementioned leak and redundancy.

Following this, we applied the derived theory to the problem of transforming %information encoded in phases 
a particular set of coherent states to a particular set of qubit states. Both sets %which 
appear in many quantum information protocols. The %first type of encoding 
considered set of coherent states are %the basis of 
so-called `phase-locked' quantum states 
%information processing systems 
(e.g. %phase-locked 
used for quantum key distribution) suitable for long-range communication, and the set of qubit states is ubiquitous in %the sphere of 
quantum computation.
For this setting, we derived the optimal transform and characterized the leak and the redundancy.
%The optimal transform is related to a family of transforms which are suitable for experimental realization in optics, 
By using beamsplitting, followed by the well-studied process of optical state truncation or `quantum scissors', %augmented by 
and an experimentally feasible amplitude re-weighing procedure, a probabilistic transform between these sets of states can be realized, albeit with sub-optimal success probability.
The success probability of %the latter family of transforms  
this procedure can however be made to asymptotically approach the optimal success probability.

An immediate application of such a transform may be in the realization of Universal Blind Quantum Computation (UBQC)~\cite{10.1109/FOCS.2009.36}, in a case where the client is restricted to producing coherent states, in contrast to the single-qubit states required by the original protocol. % and the information is encoded in the phase degree of freedom. 
%A similar setting was recently addressed 
A related procedure for this scenario was recently suggested in \cite{WCPUBQC}, where phase-randomized weak coherent states were used, %often referred to as weak coherent pulses (WCP), 
and the information was encoded in the polarization. % modes. 
This encoding the information remained essentially unitarily equivalent to the original single-qubit encoding. The question whether UBQC is possible when the client uses %similar can be achieved by using 
phase-encoded coherent states (where the unitary equivalence no longer holds) remains open. %, and a positive answer may yield interesting consequences.
Finally, the approaches developed in this paper may be applied to the task %used to combat the problem 
of amplifying coherent states \textit{truly perfectly}, which can be achieved probabilistically when the number of possible phases %which can appear 
is finite. %light 
This will be the subject of further work. %\cite{arXiv} 
% which has, to our knowledge, thus far not been fully addressed.}

\section{Acknowledgements}
%VD 
We gratefully acknowledge financial support by EPSRC grant EP/G009821/1. %, and EA \todo{fixup}
\appendix
\section{Proofs of Lemmas}
Here we give the proofs of the Lemmas which were stated in the main body of the paper. For the reasons of brevity, occasionally we will skip through technical details, and rather present the main ideas.  We begin by proving the uniformization (Lemma \ref{unif}) and symmetrization (Lemma \ref{symm})  lemmas.
\subsection*{Proof of Lemmas \ref{unif} and \ref{symm}}

\noindent \textbf{Lemma \ref{unif}} \emph{(Uniformization)
If there exists a probabilistic transformation $\mathcal{T}$ taking the states in $A$ to states in $B$ which succeeds with the probabilities $\{p_i\}_{i=1}^N$, where $A$ and $B$ are symmetric sets of states, then there exists a uniform probabilistic transform $\mathcal{T}^\prime$ taking the states in $A$ to states in $B$ which succeeds with probability
\EQ{
p = \dfrac{1}{N}\sum\limits_{i=1}^N p_i.\nonumber
}
}

\begin{figure}
\centering
\centering
\includegraphics[scale=1.3 ,viewport=130 560 550 740,clip]{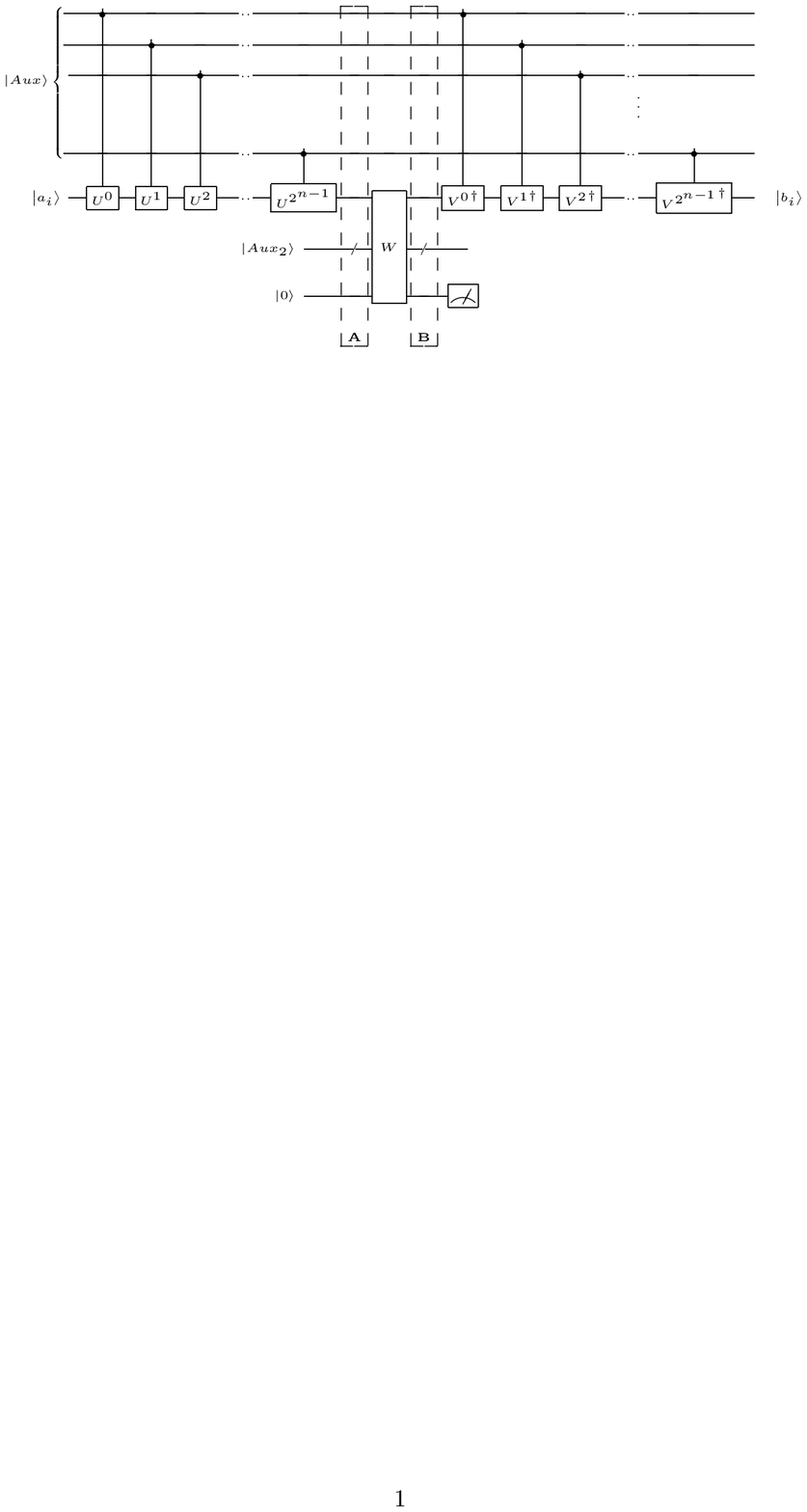}
\caption{\label{fig7} The quantum circuit by which a probabilistic transform, realized by the action of a unitary $W$ acting on an augmented Hilbert space followed by the measurement of an indicator register,  can be `uniformized'. The same circuit also serves to symmetrize the leak and the redundancy of a uniform probabilistic transform. In the proofs of lemmas we will address the states of the system above at cuts \textbf{A} and \textbf{B} denoted in this figure.}
%\begin{minipage}{\linewidth}
%\begin{flushleft}
%{\tiny
%\EQ{
%\Qcircuit @C=1em @R=1.7em {
% & \ctrl{5}& \qw &\qw & \qw & \cstick{\cdot\cdot} &  \qw &\qw &\qw &\qw  & \ctrl{5}& \qw &\qw & \qw & \cstick{\cdot\cdot} &  \qw &\qw \\
% &\qw & \ctrl{4}&\qw & \qw &\cstick{\cdot\cdot} & \qw &\qw &\qw &\qw &\qw  & \ctrl{4}&\qw & \qw &\cstick{\cdot\cdot} & \qw &\qw\\
%\lcstick{\ket{Aux}}&\qw &\qw & \ctrl{3} &  \qw &\cstick{\cdot\cdot} & \qw &\qw &\qw &\qw & \qw &\qw & \ctrl{3} &  \qw &\cstick{\cdot\cdot} & \qw &\qw\\
% & & & & & & & & &&&&&&\vdots &\\
 %&\qw & \qw & \qw & \qw & \cstick{\cdot\cdot}\ & \ctrl{1} &\qw &\qw &\qw &\qw&\qw  & \qw & \qw & \cstick{\cdot\cdot}\ & \ctrl{1} &\qw\\
%\lstick{\ket{a_i}}&\gate{U^{0}} & \gate{U^{1}} & \gate{U^{2}} & \qw & \cstick{\cdot\cdot}\ & \gate{U^{2^{n-1}}} &\qw &\multigate{2}{W}  &\qw&
%\gate{{V^{0}}^\dagger} & \gate{{V^{1}}^ \dagger} & \gate{{V^{2}}^\dagger} & \qw & \cstick{\cdot\cdot}\ & \gate{{V^{2^{n-1}}}^\dagger} &\qw & \rstick{\ket{b_i}}\\
%&\ew&\ew&\ew&\ew&\ew&\lstick{\ket{Aux_2}}&{/}\qw&\ghost{W}&{/}\qw&\qw&\ew&\ew&\ew&\ew&\ew&\ew\\
%&\ew&\ew&\ew&\ew&\ew&\lstick{\ket{0}}&\qw&\ghost{W} &\qw&\meter&\ew&\ew&\ew&\ew&\ew&\ew\\
%&\ew&\ew&\ew&\ew&\ew&\ew&\crstick{\mathbf{A}}&\ew&\crstick{\mathbf{B}}&\ew&\ew&\ew&\ew&\ew&\ew&\ew\gategroup{1}{1}{5}{5}{.7em}{\{}\gategroup{1}{8}{9}{8}{1em}{--}
%\gategroup{1}{10}{9}{10}{1em}{--}
%}
%\nonumber
%}}
%\end{
%flushleft}
%\end{minipage}
\end{figure}
%\vspace{0.5cm}
\noindent\textbf{Proof:\\}
%Consider the circuit given in Figure \ref{fig7}.
As noted, each probabilistic transform may be realized as a unitary transform acting on an augmented Hilbert space, followed by a measurement of an indicator register.
In the circuit of Figure \ref{fig7}, the transform $\mathcal{T}$ is represented by this extended unitary $W$.
As both the input and output sets of states are symmetric, there exist unitaries which sequentially shift though the states of the set, obeying the intrinsic order.
We denote these unitaries by $U$ and $V$, corresponding to the sets $A$ and $B$ respectively, and the controlled powers of these unitaries appear in the circuit.
The state $\ket{Aux}$ is pre-set to be the  uniform superposition
\EQ{
\ket{Aux} = 1/\sqrt{N} \sum\limits_{k=0}^{N-1} \ket{k},
}
where $\ket{k}$ is the $l$ qubit state of the computational basis $\ket{b_{l-1}} \otimes \cdots \otimes \ket{b_0}$, $b_j \in \{0,1\}$ for all $j$ such that $(b_{l-1}\ldots b_{0})_2 = (k)_{10}$, where the subscripts designate the base of the number representations. 

First let us show that the circuit shown performs the desired transform.
The state of the system at cut \textbf{A} in the circuit is
\EQ{
1/\sqrt{N} \sum\limits_{k=0}^{N-1} \ket{k} \ket{a_{i+k \ mod \ N}} \ket{Aux_2}\ket{0},
}
where $\ket{Aux_2}$ is some fixed auxiliary state in a sufficiently dimensional state space.
The notation we shall use corresponds to the notation used in formula \ref{ProbUnit}.
Following this, the transform $\mathcal{T}$ is applied to the register which contained the input state. The transform is explicitly realized as a unitary $W$ acting on a bigger space.
The state at cut \textbf{B} in the circuit is
\EQ{
\fl1/\sqrt{N}  \left( \sum\limits_{k=0}^{N-1}\sqrt{p_{i+k\mod \ N}} \ket{k}  \ket{b_{i+k \mod \ N}}\ket{\psi_{i+k \mod \ N}} \right) \ket{0} \nonumber\\ +1/\sqrt{N}\left(\sum\limits_{k=0}^{N-1} \sqrt{1-p_{i+k\mod \ N}} \ket{k} \ket{Fail}\ket{\phi_{i+k \ mod \ N}} \right)\ket{1} .
}
If the measurement outcome of the indicator (the last) register corresponds %ed
 to the state $\ket{0}$, then the transform has succeeded (c.f. expression \ref{ProbUnit}). From the expression above, it can be seen that this happens with probability $ p = \dfrac{1}{N}\sum_{i=1}^N p_i.$
Assume that the indicator measurement yielded the desired output. The section of the circuit after cut B undoes the controlled rotations, and the state at the end of the entire circuit is
\EQ{
 \mathcal{N}\sum_{k=0}^{N-1} \sqrt{p_{i+k\mod \ N}} \ket{k}  \ket{b_i}\ket{\psi_{i+k \mod \ N}}.
}
%\EQ{
%\ket{b_i}\otimes \mathcal{N}\sum\limits_{k=0}^{N-1} \sqrt{p_{i+k\mod \ N}} \ket{k}  \ket{\psi_{i+k \mod \ N}}.
%}
The middle register contains the desired output state, and the rest of the system contains a new leak. 
This overall procedure constitutes the new, uniformized probabilistic transform $\mathcal{T}^\prime$ from the statement of the Lemma, which succeeds with the averaged probability $p$. This proves Lemma \ref{unif}. \qed

One can verify that the new leak, generated by the `uniformized' transform described above, comprises a symmetric set of states. Using an analogous analysis, one can show that the redundancy (state generated in case of  the measurement outcome corresponding to the $\ket{1}$ state in the indicator register) is a symmetric set as well.
Now, if the extended unitary $W$ corresponds to a uniform probabilistic transform, with leak and redundancy which are not symmetric, then the extended transform of Figure \ref{fig7} will have the same success probability as $W$ itself, and the leak and redundancy will be symmetrized. Thus, the analysis above proves Lemma \ref{symm} as well. \qed
\\

\noindent \textbf{Lemma \ref{simcirc}}\emph{
A Gram matrix of kets is a circulant matrix if and only if the corresponding set of kets is symmetric. }

\noindent\textbf{Proof:\\}
Let $A = \{ \ket{a_k}\}_{k=0}^{N-1}$ be a set of kets.
We first show the necessity. If the set of kets is symmetric, then its Gram matrix is circulant.
Let $U$ be the unitary which sequentially shifts through the set of kets, obeying the intrinsic order.
Then the Gram matrix may be written as
\EQ{
G_A = \left[\bra{a_p} a_q \rangle \right]_{p=0,q=0} ^{N-1,N-1} = \left[\bra{a_0} {U ^{\dagger}} ^{p} {U } ^{q} \ket{a_0} \right]_{p=0,q=0} ^{N-1,N-1}= \nonumber \\  \left[\bra{a_0} {U } ^{q-p} \ket{a_0} \right]_{p=0,q=0} ^{N-1,N-1} = \left[\bra{a_0} {U } ^{q-p \mod\ N} \ket{a_0} \right]_{p=0,q=0} ^{N-1,N-1}.
}
It is easy to verify that the last matrix in the sequence of equalities above is circulant.

Next we show the sufficiency. If $A$ is a set of states such that its Gram matrix $G_A$ is circulant, then it is symmetric.
Since $G_A$ is a matrix of states, we have that $G_A$ allows the Cholesky decomposition, that its spectrum $\{\lambda_k\}_{k=0}^{N-1}$ is real, non-negative and sums up to $N$ (as the trace is preserved under basis change), and as it is circulant, we have that it diagonalizes in the $uDFT$ basis. Using these properties and a bit of matrix algebra, one can show that if a set of kets $\{\ket{\psi_k} \}_{k=0}^{N-1}$ has $G_A$ as a Gram matrix, then its elements can be written as%represented as follows:
\EQ{
\ket{\psi_k} = \dfrac{1}{\sqrt{N}}\sum\limits_{j=0}^{N-1} \dfrac{1}{\sqrt{\lambda_j}} e^{\frac{2 k j \pi i}{N}} \ket{b_j},
}
where the kets $\{\ket{b_k} \}_{k=0}^{N-1}$ comprise an orthonormal basis, and we define the coefficient $ \dfrac{1}{\sqrt{\lambda_j}}$ to be zero if $\lambda_j=0$.
Consider the unitary $U$, acting on %the action of which is defined on 
the $\{ \ket{b_j} \}_j$ basis as follows:
\EQ{
U\ket{b_j} = e^{\frac{2 j \pi i}{N}} \ket{b_j}.
}
By applying $U$ on the ket $\ket{\psi_k}$ we have:
\EQ{
U \ket{\psi_k} = \dfrac{1}{\sqrt{N}}\sum\limits_{j=0}^{N-1} \dfrac{1}{\sqrt{\lambda_j}} e^{\frac{2 k j \pi i}{N}} U \ket{b_j} = \nonumber \\ 
\dfrac{1}{\sqrt{N}}\sum\limits_{j=0}^{N-1} \dfrac{1}{\sqrt{\lambda_j}}e^{\frac{2 (k+1) j \pi i}{N}}  \ket{b_j} =  \ket{\psi_{k+1 \mod \ N}}.
}
Hence, the set of kets $A$ which they represent is symmetric and this proves the lemma. $\square$

The following lemmas were given in section \ref{sect}. In their proofs we shall adhere to the notation of the main body of the paper.

\noindent \textbf{Lemma \ref{lemmamulti}}\emph{
Let the amplitude $\alpha$ of the states in the set $A$, defined in equation (\ref{source}), satisfy $0 < \alpha \leq 1$. Then there exists a uniform multiprobabilistic transform with the success probability vector $(p_0, \ldots, p_{N-1})$, which takes the states from the set $A$ to the collection of target states $\{B^j\}_{j=1}^{N-1}$ \emph{and} is redundancy-free and leakless. The failure probability %the probability 
$p_0$ of this transform is equal to $\exp (-2 \alpha ^2)$.
}

\noindent\textbf{Proof:\\}
As noted in the main body of the paper, the desired transform exists if and only if 
\EQ{
G_A = p_0 \Pi^f + p_1 G_{B^1} \circ \Pi^1 + \cdots + p_{k-1} G_{B^{N-1}}\circ \Pi^{N-1}
}
holds for a vector of probabilities $(p_0, \ldots, p_{N-1})$ and for a a set of Gram matrices of states $\{\Pi^f, \Pi^1 ,\ldots ,\Pi^{N-1}\}$.
Acknowledging the requirement that this transform is leakless and redundancy-free the criterion becomes
\EQ{
G_A = p_0 \mathbf{1} + p_1 G_{B^1} + \cdots + p_{N-1} G_{B^{N-1}} \label{critmult},
}
where $ \mathbf{1}$ is a matrix with all entries being the unity.

The matrix $G_{B^j}$ can be written as \EQ{G_{B^j} = \underbrace{G_B \circ \cdots \circ G_B}_{j\ times} := G_B^{\circ j},\label{bj}} and since $G_A$, $G_B$ are circulant, and the Hadamard product of circulant matrices is circulant, $G_{B^j}$ is circulant for all $j$.
Hence all the matrices in expression (\ref{critmult}) simultaneously diagonalize in the unitary discrete Fourier transform basis so we can write this criterion in terms of vectors of eigenvalues of the corresponding matrices:
\EQ{
\lambda_{G_A} = p_0 \lambda_{\mathbf{1}} + p_1 \lambda_{G_{B^1}} + \cdots + p_{N-1} \lambda_{G_{B^{N-1}}}. \label{crit2}
}
The vector $\lambda_{\mathbf{1}}$ is the first vector of the canonical basis, that is vector with one as the first entry and zeroes elsewhere, multiplied by $N$.

It can be shown that, for any $N$, the vector of eigenvalues of $G_B$ has only the first two eigenvalues non-zero, and their value is $N/2.$
From this, using the properties given in expressions (\ref{prop1}) and (\ref{prop2}), we can see that, for $k\leq N-1$, the vector of eigenvalues of $G_B^{\circ k}$ is given by
\begin{eqnarray}
\lambda_{B^k} = \dfrac{N}{2^k}\left[ 
\left({k \atop 0}  \right),
\left({k \atop 1}  \right),\cdots,
\left({k \atop k}  \right),
0,
\cdots,
0
\right]^{T}.
\end{eqnarray}
Let $M$ be the column matrix defined by
\EQ{\label{exprr}
M = \left[N e_1 \vert \lambda_{B} \vert \lambda_{B^2} \vert \cdots \vert \lambda_{B^{N-1}} \right].
}
Then we can rewrite the condition $(\ref{crit2})$ as a system of equations,
\begin{eqnarray} \label{sys}
\lambda_{G_A} = M \ora{p}
\end{eqnarray}
where $\ora{p} = \left[p_0,\ldots, p_{N-1} \right]^T.$
Since $M$ is  upper-triangular, with non-zero element across the diagonal, it is invertible. Hence, there exists a unique vector $\ora{p}$ satisfying the system above.
The sum of the elements of a column of the matrix $M$ is $N$, so we can see (by multiplying the system $(\ref{sys})$ with the row vector $\dfrac{1}{N} \left[1, \ldots, 1 \right]$ from the left) that
$
 \sum_{i=0}^{N-1} p_i = 1,
$
as the sum of the eigenvalues of $G_A$ is $N$.

To prove the stated Lemma, we need to show that all the values $p_i$ are non-negative (for $0<\alpha \leq 1$), and that we need to show that $ p_0 = \exp(-2 \alpha^2)$ .
We begin by showing the positivity of values $p_i$, as stated, and we finish of the proof by showing  that $ p_0 = \exp(-2 \alpha^2)$ .

As noted above, the system (\ref{sys}) has a unique solution (and $M^{(-1)}$ exists), and we need to show that the solution vector comprises positive elements, i.e.
 \EQ{
M ^{(-1)}\lambda_{G_A} \label{exp1}
 }
is a vector of non-negative real numbers.
Note that the matrix $M$ can be written  as $M =M^\prime. D$, where $M^\prime$ collects all the binomial coefficients and $D$ is a diagonal matrix which appropriately assigns the weights to the columns of $M$.
The $k^{th}$ column of matrix $M^\prime$ is then given by 
\EQ{\left[ 
\left({k \atop 0}  \right),
\left({k \atop 1}  \right),\cdots,
\left({k \atop k}  \right),
0,
\cdots,
0
\right]^{T}.\nonumber}
The inverse of $M$ is then \EQ{M ^{(-1)} = D^{-1}.{M^{\prime}} ^{(-1)}. \label{Mform}} 
As the matrix $D^{-1}$ comprises only positive elements (moreover it is also diagonal), in order to show that the expression (\ref{exp1})
 is a non-negative vector, it will suffice to show that 
  \EQ{
 {M^\prime} ^{(-1)}\lambda_{G_A}
  }
is a non-negative vector. 
Let $S$ be a diagonal matrix of size $N$ of alternating signs, the first sign being positive.
Using known properties of sums of binomial coefficients, one can show that $S.M^{\prime}.S$ is the inverse of the matrix $M^\prime$. We omit the proof of this claim as the proof is technical, and the details are of no further consequence.

Now we proceed to show that each entry of the vector
\EQ{
{M^\prime}^{(-1)} \lambda_{G_A} = S.M^{\prime}.S \lambda_{G_A}
}
is non-negative, if the amplitude $\alpha$ is a positive and less or equal to unity.
Let $\lambda_i$ be the $i^{th}$ eigenvalue of the matrix $G_A$, \emph{i.e.} the $i ^{th}$ component of $\lambda_{G_A}$. Note that  the enumeration starts at zero. Then the $k ^{th}$ entry of the vector $ S.M^{\prime}.S \lambda_{G_A}$ is given by
\EQ{\label{ineqs}
(e_k)^{T} S.M^{\prime}.S \lambda_{G_A} = \sum\limits_{j=k}^{N-1} (-1)^{j+k} \left({j \atop k} \right) \lambda_{j}
}
The last entry of the vector $ S.M^{\prime}.S \lambda_{G_A}$ is the last eigenvalue of $G_A$, hence positive, so
for the expression (\ref{ineqs}) to be positive, it suffices to show that 
\EQ{
 \left({j \atop k} \right) \lambda_{j} - \left({j+1 \atop k} \right) \lambda_{j+1} \geq 0
}
for all $0 \geq k < N-1$ and $k \leq j < N-1$.
This expression simplifies to
\EQ{
 \left({j \atop k} \right) \lambda_{j} - \left({j+1 \atop k} \right) \lambda_{j+1} = \left( j \atop k \right) \left( \lambda_{j} - \dfrac{j+1}{j-k+1}\lambda_{j+1} \right).
}
Since $ \left( j \atop k \right)$ is positive, we only need to show that the following holds:
\EQ{\label{poscond}
 \lambda_{j} - \dfrac{j+1}{j-k+1}\lambda_{j+1} \geq 0.
}
In order to show this, we need to analyse the structure of the eigenvalues appearing as components of $\lambda_{G_A}$.
Recall, $\lambda_{G_A}$ was defined as the discrete Fourier transform of the first row of $G_A$. Using the expansion of coherent states in the Fock basis the $j^{th}$ eigenvalue can be given as
\EQ{
\lambda_j = \sum\limits_{l=0}^{N-1} \exp\left(-2 j l \pi i /N    \right) \sum_{r=0}^{\infty} e^{-\alpha^2} \dfrac{\alpha^{2r}}{r!} \exp (2 l r \pi i /n).
}
This can further be rearranged as follows:
\EQ{
\lambda_j&=& e^{-\alpha^2} \sum\limits_{l=0}^{N-1} \sum_{r=0}^{\infty}  \exp\left(-2 j l \pi i /N    \right) \dfrac{\alpha^{2r}}{r!} \exp (2 l r \pi i /n)\\
&=&e^{-\alpha^2}  \sum_{r=0}^{\infty} \dfrac{\alpha^{2r}}{r!}  \sum\limits_{l=0}^{N-1}  \exp (2 l (r-j) \pi i /n), \label{absconv}
}
where in order to get to expression (\ref{absconv}), we used the fact that the infinite sum above is absolutely convergent, hence allows the commuting of sums.

By the properties of sums of roots of unity, the expression  $\sum_{l=0}^{N-1}  \exp (2 l (r-j) \pi i /n)$ is equal to $n$ if $r-j$ is divisible by $N$ and zero otherwise.
Hence we get
\EQ{\label{evs}
\lambda_j = e^{-\alpha^2} N  \sum_{r=0}^{\infty} \dfrac{\alpha^{2(N r +j )}}{(N r +j)!}. 
}
The elements in the sum above appear as the summands in the Taylor expansion of $e ^ {2 \alpha}$; for $j=0$, this sum collects every $N^{th}$ summand from the Taylor series expansion, starting from the zeroth summand. For any other $j$ it collects  every $N^{th}$ summand from the Taylor series expansion, starting from the $j$-th summand.
We note that the eigenvalues above, for a fixed $N$ can be expressed in a closed form in terms of Generalized hypergeometric functions.

We set out to show that inequality (\ref{poscond}) holds. By inserting the explicit expressions for the eigenvalues we have derived, we obtain the expression 
%\red{\bf What is meant by an inequality resolving as an equality?}

\EQ{
 \lambda_{j} - \dfrac{j+1}{j-k+1}\lambda_{j+1} = \nonumber \\e^{-\alpha^2} N  \left(   \sum_{r=0}^{\infty} \dfrac{\alpha^{2(N r +j )}}{(N r +j)!}  -  \dfrac{j+1}{j-k+1}    \sum_{r=0}^{\infty} \dfrac{\alpha^{2(N r +j +1)}}{(N r +j+1)!}             \right),
}
and again by absolute convergence of the sums above we may reshuffle them and obtain
\EQ{
  e^{-\alpha^2} N   \sum_{r=0}^{\infty} \dfrac{\alpha^{2(N r +j )}}{(N r + j)!}      \left(  1 - \dfrac{j+1}{\left(j-k+1\right)} \dfrac{1}{\left(N r + j +1\right)} \alpha^2 \right).\label{finexp}
}
The expression above is positive if the expression in the last parenthesis is positive. Now we inspect the coefficient with the term $\alpha^2$ in the parenthesis,
\EQ{
\dfrac{j+1}{\left(j-k+1\right)} \dfrac{1}{\left(N r + j +1\right)}.\nonumber
}
This expression is always positive, and note that the denominator $\left(j-k+1\right)$ is greater or equal to unity, and the denominator $\left(N r + j +1\right)$ is larger or equal to $j+1$, so the entire expression is less or equal to unity.
But then for $\alpha \leq 1$ the expression (\ref{finexp}) is non-negative.

To finish the proof we need to show that $ p_0 = \exp(-2 \alpha^2)$ .
Note that $p_0 = e_0^T M^{(-1)} \lambda_{G_A}.$
Recall, $\lambda_{G_A} = DFT.(e_0^T .G_A)^T$ (i.e. the DFT of the first row of the Gram matrix of the set A is the vector of eigenvalues of $G_A$).
The exact form of $M^{-1}$ was given in expression (\ref{Mform}), and we can see that
 \EQ{e_0^T M^{(-1)} =\dfrac{1}{N} \left[1, -1, 1, \ldots, 1,-1 \right].\nonumber}
Thus, it holds that
\EQ{
p_0 = e_0^T M^{(-1)} DFT.(e_0^T .G_A)^T = \dfrac{1}{N}  \left[1, -1, 1, \ldots, 1,-1 \right].DFT.(e_0^T .G_A).\nonumber}
We can see that that  \EQ{ \left[1, -1, 1, \ldots, 1,-1 \right].DFT = N e_{N/2}, \nonumber}  as this is equivalent to adding a $\pi$ phase to each of the rows of the DFT matrix and then summing up the rows. Without the phase shift, the sum of the rows is a vector with a non-zero entry only at the first position. The phase shift corresponds to a cyclic permutation of columns by $N/2 - 1$ positions, so the sum of the rows of the permuted DFT matrix has the only non-zero entry at the $(N/2 +1)^{st}$, and this entry is $N$.  Hence we have
\EQ{p_0 = e_{N/2}.(e_0^T .G_A)^T =  \exp(-2 \alpha^2), \nonumber}
and we have proven our Lemma. $\square$

\bibliographystyle{phaip}

\end{document}